\documentclass[11pt]{article}
\pdfoutput=1

\usepackage{a4wide}
\usepackage[fleqn]{amsmath}
\usepackage{rotating}

\RequirePackage{amsmath,amssymb,amsxtra,amsthm}

\RequirePackage[T1]{fontenc}
\RequirePackage[utf8]{inputenc}

\RequirePackage{mathrsfs}
\RequirePackage{mathpazo}

\RequirePackage{setspace}
\RequirePackage{array}
\RequirePackage{booktabs}

\RequirePackage{braket}

\usepackage{cite}

\RequirePackage{colortbl}
\RequirePackage{xcolor}
\colorlet{titlerowcolor}{gray!15}
\newcommand*{\tcolrow}{\rowcolor{titlerowcolor}}

\newcommand{\be}{\begin{equation}}
\newcommand{\ee}{\end{equation}}
\newcommand{\bea}{\begin{eqnarray}}
\newcommand{\eea}{\end{eqnarray}}
\newcommand{\IR}{\mathbb{R}}

\newcommand{\IZ}{\mathbb{Z}}

\newcommand{\cF}{\mathcal{F}}

\newcommand{\cG}{\mathcal{G}}
\newcommand{\cH}{\mathcal{H}}

\newcommand{\cM}{\mathcal{M}}
\newcommand{\cW}{\mathcal{W}}

\newcommand{\cE}{\mathcal{E}}

\newcommand{\cO}{\mathcal{O}}

\newcommand{\cT}{\mathcal{T}}

\newcommand{\cZ}{\mathcal{Z}}
\newcommand{\de}{\mathrm{d}}

\newcommand{\I}{\mathrm{i}}

\newcommand{\sgn}{{\rm sgn}}

\numberwithin{equation}{section}
\numberwithin{table}{section}
\numberwithin{figure}{section}

\author{
  \begin{minipage}{0.97\linewidth}
    \vspace{1cm}
    \begin{center}
      \begin{small}
        \textbf{Carlo Angelantonj}$^{1,2}$, \textbf{Ioannis Florakis}$^{3,4}$ and  \textbf{Boris Pioline}$^{2,5}$
     \end{small}
    \end{center}
    \vspace{.3cm} \hspace{1.3cm}\begin{minipage}{.75\linewidth}
      {\it \begin{footnotesize}
          \begin{itemize}
          \item[${}^1$] Dipartimento di Fisica, Universit\`a di Torino, and INFN Sezione di Torino
          \\
            Via P. Giuria 1, 10125 Torino, Italy
          \item[${}^2$] CERN Dep PH-TH, 1211 Geneva 23, Switzerland
          \item[${}^3$] Arnold Sommerfeld Center for Theoretical Physics\\
            Fakult\"at f\"ur Physik, Ludwig-Maximilians-Universit\"at M\"unchen\\
           Theresienstr. 37, 80333 M\"unchen, Germany
          \item[${}^4$] Max-Planck-Institut f\"{u}r Physik,\\
	    Werner-Heisenberg-Institut,
              80805 M\"{u}nchen, Germany
         \item[${}^5$] Laboratoire de Physique Th\'eorique et Hautes Energies, CNRS UMR 7589,
         \\
         Universit\'e Pierre et Marie Curie - Paris 6, 4 place Jussieu,
         75252 Paris cedex 05, France
          \end{itemize}
        \end{footnotesize}}
    \end{minipage}
    \vspace{1cm}
  \end{minipage}
}

\date{}

\title{\vspace{3cm}
  \begin{LARGE}
    \textbf{One-Loop BPS amplitudes 
    as BPS-state sums}
  \end{LARGE}
}

\begin{document}

\begin{titlepage}
  \maketitle
  \thispagestyle{empty}

  \vspace{-14cm}
  \begin{flushright}
    CERN-PH-TH/2012-061\\  
    DFTT 04/2012\\
    LMU-ASC 15/12\\
    MPP-2012-52\\
    arXiv:1203.0566v4 
   \end{flushright}

  \vspace{11cm}

  \begin{center}
    \textsc{Abstract}\\
  \end{center}
Recently, we  introduced a new procedure for computing a class of one-loop BPS-saturated amplitudes in String Theory, which expresses them as a  sum of one-loop contributions of all perturbative BPS states in a manifestly T-duality invariant fashion. 
In this paper, we extend  this procedure to {\em all} 
BPS-saturated amplitudes of the form $\int_{\cF} \varGamma_{d+k,d}\, \varPhi$, with $\varPhi$ 
being a weak (almost)  holomorphic modular form of weight $-k/2$. We use the fact that any such $\varPhi$ can be expressed as a linear combination of  certain absolutely convergent Poincar\'e series, against
which the fundamental domain $\cF$ can be unfolded. The resulting BPS-state
sum neatly exhibits  the singularities  of the amplitude at points of  gauge symmetry enhancement, in a chamber-independent fashion. We illustrate our method with concrete examples of interest in heterotic string compactifications.
\vfill
{\small
\begin{itemize}
\item[E-mail:] {\tt carlo.angelantonj@unito.it}\\ {\tt florakis@mppmu.mpg.de}\\
{\tt boris.pioline@cern.ch}
\end{itemize}
}
\vfill

\end{titlepage}

\setstretch{1.1}

\tableofcontents


\section{Introduction}

Scattering amplitudes in closed string theory involve, at $h$-th order in perturbation  
theory, an integral over the moduli space of conformal structures on genus $h$ closed Riemann surfaces.  The torus amplitude (corresponding to $h=1$) is particularly relevant, as it encodes the perturbative spectrum of excitations. Moreover, for special choices of vacua and of external states, corresponding to a special class of $F$-term interactions in the low energy effective action, the torus contribution exhausts the perturbative series, and thus can serve as a basis for quantitative tests of string dualities (see e.g. \cite{Kiritsis:1999ss} and references therein).

The moduli space of conformal metrics on the torus is the 
Poincar\'e upper half plane ${\mathbb H}$, parameterised by the complex structure parameter $\tau=\tau_1+\I \tau_2$, modulo the action of the modular group 
${\rm SL} (2,\IZ)$. After performing the path integral over the world-sheet 
fields and over the location of the vertex-operator insertions, the relevant amplitude is then expressed as a modular integral
\begin{equation}
\label{intgen}
\int_{\mathcal F} \de \mu\, {\mathcal A} (\tau_1 , \tau_2 )\ ,
\end{equation}
where ${\mathcal F} = \{ \tau \in \mathbb{H} \, |\, -\frac{1}{2} \le \tau_1 < \frac{1}{2}, |\tau|\ge 1\}$
is the standard fundamental domain, $\de\mu = \tau_2^{-2}\, 
\de\tau_1\,\de\tau_2$ is the ${\rm SL} (2,\mathbb{Z})$-invariant integration measure, and ${\mathcal A}$ 
is a modular-invariant function  whose precise expression depends on the problem at hand. 
With this choice of integration domain,  the imaginary part  $\tau_2$ can be identified with Schwinger's proper time, while the real part $\tau_1$ is the Lagrange multiplier  imposing the level-matching condition. 
Part of the difficulty in evaluating integrals of the form \eqref{intgen} is the unwieldy
shape of  $\cF$, which intertwines the integrals over $\tau_1$ and $\tau_2$. 

Depending on the function ${\mathcal A}(\tau_1 , \tau_2 )$ methods have been devised to overcome this problem. If ${\mathcal A}$ is a weak almost  holomorphic function\footnote{By {\em weak almost  holomorphic} we mean  an element in the graded 
polynomial ring generated by the holomorphic Eisenstein series $E_4$ and $E_6$, the almost holomorphic Eisenstein series $\hat E_2$ and the inverse of the discriminant $1/\varDelta$. Our notations for Eisenstein series and other modular forms are collected in Appendix \ref{sec_notmod}. The adverb {\em weak} refers to the fact that the only singularity is, at most, a finite order pole at the cusp $q=0$.}  of $\tau$ (or, alternatively, an anti-holomor\-phic function), the surface integral over ${\mathcal F}$ can be reduced by Stokes' theorem to a line-integral over its boundary $\partial{\mathcal F}$ that can be explicitly computed  \cite{Lerche:1987qk}. On the contrary, if ${\mathcal A}$ is a genuine non-holomorphic function, as is the case for the one-loop partition function of closed-oriented strings, no useful method is known to evaluate the integral, but one can use the Rankin-Selberg-Zagier transform \cite{MR656029} to connect the integral to the graded sum of physical degrees of freedom \cite{Kutasov:1990sv, Angelantonj:2010ic, Cardella:2010bq, Cardella:2008nz}. A frequently encountered intermediate case is that of modular integrals of the form
\begin{equation}
\int_{\mathcal F}\de\mu\,  \varGamma_{d+k,d} (G,B,Y ; \tau_1 , \tau_2 ) \, \varPhi ( \tau) \,, 
\label{modintegral}
\ee
where 
\be
 \varGamma_{d+k,d}(G,B,Y; \tau_1 , \tau_2) \equiv \tau_2^{d/2} \, 
 \sum_{p_{\rm L}, p_{\rm R}} \, q^{\frac14 p_L^2}\, \bar q^{\frac14 p_R^2}
 \label{defGamma}
\end{equation}
is the partition function of the Narain lattice of Lorentzian signature $(d+k,d)$, $G,\, B, \, Y$ parameterise the Narain moduli space ${\rm SO} (d+k,d)/{\rm SO} (d+k) \times {\rm SO} (d)$, 
and $\varPhi (\tau )$ is a weak almost  holomorphic modular form of negative weight $w=-k/2$, which we shall
refer to as the {\em elliptic genus}. Such
integrals occur in particular in one-loop corrections to certain BPS-saturated couplings
in the low energy effective action of heterotic or type II superstrings.

The traditional approach in the physics literature for computing modular integrals of the 
form \eqref{modintegral} has been  the {\em orbit method}, which proceeds by unfolding
the integration domain $\cF$ against the lattice partition function $\varGamma_{d+k,d}$ \cite{McClain:1986id,O'Brien:1987pn,Dixon:1990pc,Mayr:1993mq,Harvey:1995fq,Harvey:1996gc, Bachas:1997mc, stst1, stst2, Kiritsis:1997hf, Kiritsis:1997em, Marino:1998pg}.  While this procedure yields an infinite series
expansion which is useful in certain limits in Narain moduli space, it does not make 
manifest the  invariance under the T-duality group ${\rm O}(d+k,d,\IZ)$ of the Narain lattice, nor does it clearly display the singularities of the amplitude at points of  gauge symmetry enhancement. 

In  \cite{Angelantonj:2011br} we proposed a new method for  dealing with
modular integrals of the form \eqref{modintegral},
which relies on representing 
the elliptic genus $\varPhi$ as a Poincar\'e series, and on unfolding  the integration domain against 
{\it it} rather than against the 
lattice partition function. The advantage of this method is that T-duality  remains manifest at all steps,
and the result is valid in all chambers in Narain moduli space, unlike the conventional approach\footnote{See for instance \cite{Harvey:1995fq} for a detailed discussion on chamber dependence of the traditional unfolding method.}. Moreover, the amplitude is expressed as a sum over all BPS states 
in the spectrum, thus generalising the constrained Eisenstein series
constructed in \cite{Obers:1999um}. \footnote{BPS states sums have appeared in earlier works \cite{Ooguri:1991fp, Ferrara:1991uz, LopesCardoso:1995qa, LopesCardoso:1994ik, Lerche:1999ju}. In our approach these BPS sums follow directly from unfolding the fundamental domain against the elliptic genus, without any further assumption.} Finally, the singularities of the amplitude at points of enhanced  gauge symmetry can be immediately read-off from the contributions of those BPS states which become massless. 

The main difficulty in implementing this strategy   
is due to the fact that the standard Poincar\'e series representation of a weak holomorphic
modular form of weight $w\leq 0$ \cite{0306.30023,0695.10021,Manschot:2007ha}
is only conditionally convergent, and therefore unsuited for unfolding. In \cite{Angelantonj:2011br} we  attempted to circumvent this problem by considering a class of non-holomorphic Poincar\'e series $E(s,\kappa,w)$ that provide a natural regularisation of the modular forms of interest
by inserting a Kronecker-type convergence factor  $\tau_2^{s-w/2}$ 
in the standard sum over images.  Therefore, the resulting Poincar\'e series, originally studied in 
\cite{0142.33903},  converges absolutely for $\Re(s)>1$, and  the modular integral $\int_{\cF} \varGamma_{d,d}\, E(s,\kappa,w)$ can  be 
computed by unfolding $\cF$ against it, at least for large $s$. 
The result should then be  analytically continued to the desired value $s=\tfrac{w}{2}$, where $E(s,\kappa,w)$ becomes formally a
holomorphic function of $\tau$. This procedure would then allow to  compute the modular integral 
\eqref{modintegral} for any $\varPhi$ which can be expressed as a linear combination
of such $E(\tfrac{w}{2},\kappa,w)$'s, at least in principle. However, this strategy turned out to be quite difficult in practice, since this analytic continuation depends on the notoriously subtle  analytic properties  of the Kloosterman-Selberg zeta function which appears in the Fourier expansion 
of $E(s,\kappa,w)$.
That is the reason why the analysis \cite{Angelantonj:2011br}
was restricted to the case of  zero modular weight, where the analytic continuation is fully under control. 

In the present work, we overcome these difficulties by employing a different class of non-analytic Poincar\'e series introduced
in the mathematics literature by Niebur \cite{0288.10010} and Hejhal \cite{0543.10020}
and studied more recently  by Bruinier, Ono and Bringmann \cite{1004.11021,BruinierOno,1154.11015,OnoMockDelta}. Similarly to the Selberg-Poincar\'e series $E(s,\kappa,w)$,
the Niebur-Poincar\'e series $\cF(s,\kappa,w)$ converges absolutely for $\Re(s)>1$, 
and formally becomes holomorphic in $\tau$ at the point $s=\tfrac{w}{2}$. 
However,  the Niebur-Poincar\'e series  can be specialised to the other 
interesting value  $s=1-\tfrac{w}{2}$, which lies  inside the domain of absolute convergence when the weight $w$ is negative. Although at this value  $\cF(s,\kappa,w)$ belongs to the more general class of  weak harmonic Maass forms\footnote{A harmonic Maass form  is 
an eigenmode of the weight-$w$ Laplacian on $\mathbb{H}$ with the same eigenvalue as weak 
holomorphic modular forms. The positive frequency part of a weak harmonic Maass 
form is sometimes known as a Mock modular form. See Section \ref{defMaass} for a more precise definition of weak harmonic Maass forms.}, that are typically non-holomorphic functions of $\tau$, 
it has the important property that any linear combination of $\cF(1-\tfrac{w}{2},\kappa,w)$, whose coefficients are determined by
the principal part of a given weak holomorphic modular form $\varPhi$, is actually
a weak holomorphic modular form, and equals $\varPhi$ itself. Therefore, {\em given any  weak holomorphic modular form $\varPhi$, the integral \eqref{modintegral}
can be computed by decomposing $\varPhi$ into a sum of Niebur-Poincar\'e series,
and by unfolding each of them against the integration domain}.
 Moreover, the same strategy works also for weak almost holomorphic modular forms ({\em i.e.} involving  powers of $\hat E_2$), where now one has to specialise the Niebur-Poincar\'e series to the values $s=1-\tfrac{w}{2}+n$, with $n$ a non-negative integer.

The outline of this work is as follows. In Section \ref{sec_Niebur}, we introduce the 
Niebur-Poincar\'e series $\cF(s,\kappa,w)$, discuss their main properties,
present their Fourier coefficients and identify their limiting values 
at $s=1-\tfrac{w}{2}+n$. We conclude the section by showing the important result that {\em any}  weak almost holomorphic modular form can be represented as a linear combination of them.
In Section \ref{integrals} we evaluate
the modular integral $\int_{\cF} \varGamma_{d+k,d} \, \cF(s,\kappa,-\frac{k}{2})$  in terms of certain BPS-state sums and discuss their singularity structure.  In Section \ref{SecExamples}, we use this result to compute a sample of modular integrals of physical interest
of the form \eqref{modintegral}. In Appendix \ref{sec_not}, we define
our notation for modular forms, we collect various definitions and properties of Whittaker and hypergeometric functions, and we introduce the Kloosterman sums and the Kloosterman-Selberg zeta function. Finally, in Appendix \ref{sec_Selberg} we briefly discuss the relation between
the Selberg- and Niebur-Poincar\'e series, and between the ``shifted constrained'' Epstein zeta series and the above BPS-state sums.  The reader interested only in physics applications may skip Section \ref{niebur} and proceed directly to Section \ref{integrals}, which begins with an executive summary of the main properties of $\cF(s,\kappa,w)$.

\bigskip

{\em Note.} After having obtained most of the results in this paper, we became aware of ref. \cite{1004.11021} where similar computations have been performed for general even lattices of signature $(d+k,d)$ with $d=0,1,2$, in particular reproducing Borcherds' automorphic
products for $d=2$ \cite{0919.11036}. Unlike  \cite{1004.11021}, we restrict the analysis to even self-dual lattices (with $k=0\ {\rm mod}\ 8$), which suffices for our physics applications, but we  allow for almost holomorphic modular forms and  arbitrary dimension $d$.

\section{Niebur-Poincar\'e series and almost holomorphic modular forms \label{sec_Niebur}}
\label{niebur}

In this section, we introduce  the Niebur-Poincar\'e series $\cF(s,\kappa,w)$, a modular invariant regularisation of the na\"{\i}ve Poincar\'e series of negative weight. We present its Fourier expansion for general values of $s$, and analyse its limit as $s \to 1-\tfrac{w}{2}+n$ where $n$ is any non-negative integer. 
We explain how to represent any weak almost  holomorphic modular form of negative weight  as a suitable  linear combinations of such Poincar\'e series.

\subsection{Various Poincar\'e series}

In order to motivate the construction of the Niebur-Poincar\'e series, let us start with a brief overview of Poincar\'e series in general. Let $w$ be an even integer\footnote{In this paper we shall restrict to the case of even weight $w$ in order to avoid complications with non-trivial multiplier systems, though the construction can be generalised to half-integer weights.} and $f$ a function on the Poincar\'e upper half plane ${\mathbb H}$. The action of an element $\gamma={\scriptsize \begin{pmatrix} a & b \\ c & d \end{pmatrix}} \in \varGamma= {\rm SL} (2,\IZ)$ on $f$ is given by the Petersson slash operator 
\begin{equation}
\left( f\vert_{w}\gamma \right)(\tau) = (c\tau+d)^{-w}\, f(\gamma\cdot\tau)\ ,\qquad
\gamma\cdot\tau = \frac{a \tau+b}{c\tau+d}\ .
\end{equation}
If $f$ is invariant under $\varGamma_\infty = {\scriptsize \begin{pmatrix} 1 & \star \\ 0 & 1 \end{pmatrix}}\subset\varGamma$, the Poincar\'e series of  {\em seed} $f$ and weight $w$
\begin{equation}
\label{Pseed}
P(f,w;\tau) \equiv P(f,w) = 
\tfrac{1}{2} \, \sum_{\gamma\in \varGamma_\infty\backslash \varGamma} f\vert_{w}\gamma
\end{equation}
defines an automorphic form of weight $w$ on $\mathbb{H}$, which is absolutely convergent provided 
$f(\tau) \ll \tau_2^{1-\frac{w}{2}}$ as $\tau_2\to 0$. As an example, the choice $f(\tau)=q^{-\kappa}$ with $w>2$ leads to the usual holomorphic  Poincar\'e series
 \begin{equation}
\label{defPgen}
P (\kappa ,w) =  \tfrac{1}{2} \sum_{(c,d)=1} 
(c\tau+d)^{-w}\,
\, e^{-2\pi\I\kappa\, \frac{a\tau+b}{c\tau+d}}\, ,
\end{equation}
where the pair $(a,b)$ is determined modulo $(c,d)$ by the condition $ad-bc=1$. Depending on the value of $\kappa$, Eq. \eqref{defPgen} describes different types of modular forms. 
For $\kappa=0$, $P(\kappa,w)$ is actually an Eisenstein series, while for $\kappa\leq -1$ it is a cusp  form, and must therefore vanish if $2<w<12$, an observation that will be important  later. For $\kappa>0$, Eq. \eqref{defPgen} represents instead a weak holomorphic modular form with a pole of order $\kappa$ at $q=0$, $P(\kappa,w)=q^{-\kappa}+{\mathcal O} (q)$.

For $w\leq 2$,  the Poincar\'e series \eqref{defPgen} is divergent and thus needs to be regularised. One possible regularisation scheme, introduced in the mathematical literature in  \cite{0306.30023,0695.10021} and
discussed in the physics literature in \cite{Manschot:2007ha}, is to consider  the convergent sum 
\begin{equation}
\label{PoincaHol}
P(\kappa,w) = \tfrac12\, \lim_{K\to\infty} \sum_{|c|\leq K}\quad 
\sum_{|d|<K; (c,d)=1} (c\tau+d)^{-w} \,  e^{2\pi\I \kappa \,\frac{a\tau+b}{c\tau+d}} 
\, R\left( \frac{2\pi\I|\kappa |}{c(c\tau+d)}\right) \ ,
\end{equation}
where $R$ is a specific regulating factor such that $R(x)\sim x^{1-w}/\varGamma(2-w)$ as $x\to 0$ and approaches 1  as $x\to\infty$. While this regularisation preserves holomorphicity, it does not  necessarily produce a modular form\footnote{The holomorphic Poincar\'e series \eqref{PoincaHol} is in general an Eichler integral, {\em i.e.} a function $F(\tau)$ which satisfies $F(\tau)- (F|_w \gamma )(\tau) =r_\gamma(\tau)$ where $r_\gamma$ is a polynomial of degree $-w$ in $\tau$, whose coefficients depend on $a,b,c,d$. We shall comment in Section \ref{defMaass} on the modular completion of $P(\kappa,w)$.}, except for small $|w|$ where the modular anomaly can be shown to vanish. Moreover, the convergence of \eqref{PoincaHol} is  conditional, which makes it unsuitable for the unfolding procedure.

Another option, introduced  by Selberg  \cite{0142.33903} and considered in our previous work \cite{Angelantonj:2011br}, is to jettison  holomorphicity and introduce a convergence factor {\em \`a la} Kronecker, thus considering the Poincar\'e-series
 \begin{equation}
\label{defEgen}
E (s,\kappa ,w)\equiv  \tfrac{1}{2} \sum_{(c,d)=1} 
\frac{\tau_2^{s-\frac{w}{2}}}{|c\tau+d|^{2s-w}}\, (c\tau+d)^{-w}\, 
e^{-2\pi\I\kappa\, \frac{a\tau+b}{c\tau+d}}\, 
\end{equation}
associated to the seed $f(\tau)=\tau_2^{s-\frac{w}{2}}\, q^{-\kappa}$. We shall refer to \eqref{defEgen} as the Selberg-Poincar\'e series. The series \eqref{defEgen} converges absolutely for $\Re(s)>1$ and becomes formally holomorphic at $s=\tfrac{w}{2}$. However, for $w\leq 2$ this  value lies outside the convergence domain, and the analytic continuation to $s=\tfrac{w}{2}$  depends on the analytic properties of the Kloosterman-Selberg Zeta function, defined in Appendix \ref{sec_Selberg}, which are notoriously subtle. In particular this analytic continuation generally leads to holomorphic anomalies. For this reason, in \cite{Angelantonj:2011br} we restricted the analysis to the case $w=0$, where the analytic continuation is under control. Another drawback of the Selberg-Poincar\'e series \eqref{defEgen} is that it fails to be an eigenmode of the  Laplacian on $\mathbb{H}$, rather it satisfies \cite{0507.10029}
\begin{equation}
\label{laplEskw}
\left[\Delta_w + \tfrac{1}{2}\,  s(1-s) +\tfrac{1}{8}\, w(w+2)\right] \, E(s,\kappa,w) = 
2\pi\kappa\, ({s-\tfrac{w}{2})}\, 
E(s+1,\kappa,w)\, ,
\end{equation}
where $\Delta_w$ is the weight-$w$ hyperbolic Laplacian defined in \eqref{Deltaw}. Since $E(s+1,\kappa,w)$ may in general have a pole at $s=\tfrac{w}{2}$, the analytic continuation of $E(s,\kappa,w)$ to this value is not even guaranteed to be harmonic.

To circumvent these problems,  following \cite{0288.10010,0543.10020,1004.11021}
we introduce a different regularisation of the Poincar\'e series \eqref{defPgen} for negative weight, which is both  modular invariant and  annihilated by the operator on the {\em l.h.s.} of \eqref{laplEskw}. Namely, we choose the seed  in \eqref{Pseed} to be $f(\tau)=\cM_{s,w}(-\kappa\tau_2)\, e^{-2\pi\I\kappa\tau_1}$ 
where 
\begin{equation}
\cM_{s,w}(y) = |4\pi y|^{-\frac{w}{2}}\, M_{\frac{w}{2}\sgn(y), s-\frac12}
\left(4\pi |y| \right)
\label{curlyM}
\end{equation}
is expressed in terms of the  Whittaker function\footnote{For a definition of Whittaker functions and 
some of their properties see Appendix \ref{WhittakerApp}.} $M_{\lambda,\mu}(z)$.
We thus define the Niebur-Poincar\'e series
\begin{equation}
\begin{split}
\label{Fskw}
\cF(s,\kappa,w) =&\tfrac12 \sum_{\gamma\in \varGamma_\infty\backslash \varGamma} \,
\cM_{s,w}(-\kappa\tau_2)\, e^{-2\pi\I\kappa\tau_1}\, \vert_w\, \gamma \\
=& \tfrac{1}{2} \sum_{(c,d)=1} (c\tau + d)^{-w}\, {\mathcal M}_{s,w} \left(- \frac{\kappa\, \tau_2}{|c\tau + d|^2} \right) \, \exp \left\{ -2\I\pi \kappa \left( \frac{a}{c} -\frac{c\tau_1+d}{c|c\tau + d|^2}\right)\right\} \,.
\end{split}
\end{equation}
Since $\cM_{s,w}(y)\sim |4\pi y|^{s-\frac{w}{2}}$ as $y\to 0$, 
 Eq. \eqref{Fskw} converges absolutely for $\Re(s)>1$, independently of $w$ and $\kappa$.  Moreover, for $\kappa>0$, the case of main interest in this work, the seed behaves as
\begin{equation}
\label{limMcusp}
\cM_{s,w}(-\kappa \tau_2)\,e^{-2\pi\I\kappa\tau_1} 
 \sim \frac{\varGamma(2s)}{\varGamma(s+\frac{w}{2})}\, q^{-\kappa}\qquad \mbox{as}\quad
 \tau_2\to\infty\ ,
\end{equation}
so that $\cF(s,\kappa,w)$ can indeed be viewed as a regulated version of the na\"{\i}ve Poincar\'e series $P(q^{-\kappa},w)$, up to an overall normalisation. By construction it is an eigenmode of the weight-$w$ Laplacian on $\mathbb{H}$,
\begin{equation}
\label{laplFskw}
\left[\Delta_w + \tfrac{1}{2}\, s(1-s) +\tfrac{1}{8}\, w(w+2)\right] \, \cF(s,\kappa,w) = 0 \,,
\end{equation}
for all values of $s,\kappa,w$. 
We shall denote by $\cH(s, w) = \cH(1-s,w)$ the space of real-analytic solutions to \eqref{laplFskw} which transform with modular weight $w$ under $\varGamma$. 

The raising and lowering operators $D_w$, $\bar D_w$ defined in \eqref{modularderiv},  map $\cH(s,w)$ into $\cH(s, w\pm 2)$, and have a simple action on the Niebur-Poincar\'e series 
\begin{equation}
\label{DwF}
\begin{split}
D_w\cdot \cF(s,\kappa,w) &= 2\kappa\, (s+\tfrac{w}{2})\,   \cF(s,\kappa,w+2)\,,
\\
\bar D_w \cdot \cF(s,\kappa,w) &= \frac{1}{8\kappa} (s-\tfrac{w}{2})\,   \cF (s,\kappa,w-2)\,.
\end{split}
\end{equation}
Furthermore,   under the action of the Hecke operator \eqref{defHecke} $\cF(s,\kappa,w)$ transforms as
\begin{equation}
\label{HeckeF}
T_{\kappa'}\cdot   \cF(s,\kappa,w) =  
\sum_{d|(\kappa,\kappa')} d^{1-w}\, \cF(s,\kappa\kappa'/d^2,w)\ .
\end{equation}
In particular, setting $\kappa=1$, the series $\cF(s,\kappa',w)$ is obtained
by acting with $T_{\kappa'}$ on $\cF(s,1,w)$.

While the Poincar\'e series \eqref{Fskw} converges absolutely only for $\Re(s)>1$, it is 
known to have a meromorphic continuation to the complex $s$-plane, holomorphic 
in the region $\Re(s)>\frac{1}{2}$ \cite{0288.10010,0352.30012}, but  with poles on the lines $s\in \frac{1}{2} + \I\IR$ and $s\in \frac{1}{4} + \I\IR$. Moreover, the `completed' series 
\begin{equation}
\cF^\star(s,\kappa,w) = 
\frac{\varGamma (1-2s)}{\varGamma (1-s+\frac{w}{2}\,{\rm sgn} (\kappa))} \cF(s,\kappa,w)
\end{equation}
is known to be odd under $s\mapsto 1-s$, up to an additive contribution proportional
to the non-holomorphic Eisenstein series $E(s,0,w)$ \cite{0288.10010,0352.30012}.
 In this work
however we shall only consider $\cF(s,\kappa,w)$ in its domain of convergence $\Re(s)>1$,
except for $w=0$ where we allow $s=1$.

\subsection{Fourier expansion of the Niebur-Poincar\'e series}

The Fourier expansion of $\cF(s,\kappa,w)$ can be obtained following the standard procedure of extracting the contribution from $c=0, d=1$, setting $d=d'+m c$ in the remaining sum, and Poisson resumming over $m$. The result is \cite{1004.11021,1154.11015}
\begin{equation}
\label{FskwF}
\cF(s,\kappa,w)=\cM_{s,w}(-\kappa \tau_2)\, e^{-2\pi\I\kappa\tau_1}
+ \sum_{m\in\IZ} \, \tilde\cF_m(s,\kappa,w) \, e^{2\pi\I m\tau_1}\,,
\end{equation}
where, for zero frequency,
\begin{equation}
\tilde\cF_{0}(s,\kappa,w) =\frac{2^{2-w}\, \I^{-w}\, \pi^{1+s-\frac{w}{2}}\, |\kappa|^{s-\frac{w}{2}}\, \varGamma (2s-1)\, \sigma_{1-2s} (\kappa )}{\varGamma(s-\frac{w}{2}) \, \varGamma(s+\frac{w}{2})\, \zeta(2s)}
\, \tau_2^{1-s-\frac{w}{2}}\,,
\label{Fzero}  
\end{equation}
while for non-vanishing integer frequencies\footnote{Note that $\tilde\cF_{-\kappa<0}$  does {\it not} include the contribution from the first term in \eqref{FskwF}.}
\begin{equation}
\tilde\cF_{m}(s,\kappa,w)  =\frac{4\pi\, |\kappa|\, \I^{-w}\, \varGamma(2s)}{ \varGamma ( s+\frac{w}{2} \, {\rm sgn}(m))}\, 
 \left|  \frac{m}{\kappa}\right|^{\frac{w}{2}} \, 
 \cZ_s(m,-\kappa)\, \cW_{s,w}(m\tau_2)\,.
\label{FourierFBO} 
\end{equation}
In these expressions,  $\sigma_s (k)= \sum_{d|k} d^s$ is the divisor function and  $\cZ_s(m,-\kappa)$ is the Kloosterman-Selberg zeta function \eqref{Ziv}, a number-theoretical function which plays a central r\^ole in the theory of Poincar\'e series. The function $\cW_{s,w}$ is expressed in terms of the Whittaker $W$-function as 
\begin{equation}
\label{defW}
\cW_{s,w}(y) = |4\pi y|^{-\frac{w}{2}}\, W_{\frac{w}{2}\sgn(y), s-\frac12}
\left(4\pi |y| \right) 
\,,
\end{equation}
and is determined uniquely by the requirement that  $\cW_{s,w}(n\tau_2)\,  e^{2\pi\I m\tau_1}$  be annihilated by the Laplace operator on the {\em l.h.s.} of \eqref{laplFskw}, and be exponentially suppressed as $\tau_2\to\infty$. 

Using the properties \eqref{DwM} and \eqref{DwW},  it is straightforward to check that all Fourier modes transform according to \eqref{DwF} under the raising and lowering operators $D_w, \bar D_w$. Moreover, using the action \eqref{Heckemodes} of the Hecke operators on the Fourier coefficients, and the Selberg identity \eqref{SelbergId} satisfied by the Kloosterman sums, one can show that
\begin{equation}
\label{HeckeF1}
T_\kappa\cdot \cF(s,1,w) = \cF(s,\kappa,w) \,.
\end{equation}
Eq. \eqref{HeckeF} follows then from this equation and from the Hecke algebra \eqref{HeckeAlg}.

\subsection{Harmonic Maass forms from Niebur-Poincar\'e series} 
\label{defMaass}

Let us focus on the Niebur-Poincar\'e series $\cF(s,\kappa,w)$ at the point $s = 1-\tfrac{w}{2}$. To motivate this value,  we recall that any weak holomorphic modular form is an eigenmode of $\Delta_w$ with eigenvalue $-\tfrac{w}{2}$, and therefore  belongs to $\cH(s,w)$ for $s=1-\tfrac{w}{2}$ (or equivalently, $s=\tfrac{w}{2}$). However, weak holomorphic modular forms are not the only eigenmodes of $\Delta_w$ with this eigenvalue. In fact, the space  $\cH(1-\tfrac{w}{2},w)$
is  known as the space of {\em weak harmonic Maass forms} of weight $w$, of which weak holomorphic modular forms are only a proper subspace. 

The Fourier expansion of a general weak harmonic Maass form $\varPhi$ of weight $w$ is given by \cite{1088.11030}
\begin{equation}
\label{genharm}
\varPhi = \sum_{m=-\infty}^{-1} (-m)^{w-1}\, 
\bar b_{-m}\, \varGamma(1-w, -4\pi m\tau_2)\, q^{m} +
\frac{\bar b_0\, (4\pi\tau_2)^{1-w}}{w-1}
+ \sum_{m=-\kappa}^{\infty} a_m\, q^m \ ,
\end{equation} 
where $\varGamma(s,x)$ is the incomplete Gamma function and $a_m,b_m$ are coefficients constrained by modular invariance. As a result, a generic weak harmonic Maass form has an infinite number of negative frequency components, which are non-holomorphic functions of $\tau$. A harmonic Maass form splits into the sum $\varPhi=\varPhi_a+\varPhi_b$ of a holomorphic part $\varPhi_{a}=\sum_{m=-\kappa}^{\infty} a_m\, q^m$, sometimes called a Mock modular form, and a non-holomorphic part $\varPhi_b$. The non-holomorphic and holomorphic parts  can be extracted using the lowering operator $\bar D_w$ and the iterated raising operator $D_w^{1-w}$. 
Indeed, 
\begin{itemize}
\item the operator $\bar D_w$ annihilates the holomorphic part, and produces, up to powers  of $\tau_2$, the complex conjugate of a holomorphic modular form $\varPsi$ of weight $2-w$, 
\begin{equation}
\bar D_w\cdot \varPhi =\bar D_w\cdot \varPhi_b =- 2^{1-2w}\, (\pi \tau_2)^{2-w} \, \overline{\varPsi}\ ,\qquad
\varPsi (\tau)=
\sum_{m=0}^{\infty}\,  b_m  \, q^m \,,
\end{equation}
sometimes known as the {\em shadow}.
\item
the iterated raising operator $D_w^{1-w}$, also known in the physics literature as the Farey transform  \cite{Dijkgraaf:2000fq}, annihilates the non-holomorphic part, and produces a weak holomorphic modular form $\varXi$ of weight $2-w$,
\begin{equation}
D_w^{1-w}\cdot \varPhi = D_w^{1-w}\cdot \varPhi_a = \varXi\ ,\qquad \varXi\equiv \sum_{m=-\kappa}^{\infty} \, (-2m)^{1-w}\, a_m \, q^m \ ,
\end{equation}
that we shall call the {\em ghost}. The ghost encodes the holomorphic part of the harmonic Maass form (modulo an additive constant)\footnote{Notice that the ghost is only defined for integer weight $w$, unlike the shadow, which extends to the case of half-integer weight Mock theta series. }.
\end{itemize}

Returning to the Niebur-Poincar\'e series, we see that by construction the series  $\cF(s,\kappa,w)$ at the special point $s=1-\tfrac{w}{2}$ --- which, for $w<0$, {\em  belongs to the convergence domain} --- is a weak harmonic Maass form of weight $w$. Indeed, using \eqref{limM3} we find that its Fourier expansion \eqref{FskwF} reduces to 
\begin{equation}
\cF(1-\tfrac{w}{2},\kappa,w) =\cM_{1-\frac{w}{2},w}(-\kappa\tau_2) \, e^{-2\pi\I\kappa\tau_1} 
+ \sum_{m\in\IZ} \, \tilde\cF_m(1-\tfrac{w}{2},\kappa,w)  \, e^{2\I \pi m \tau_1}\,,
\end{equation}
where the seed simplifies to a finite sum 
\begin{equation}
\label{Mharm}
\begin{split}
{\mathcal M}_{1-\frac{w}{2},w}(-\kappa\tau_2) \, e^{-2\pi\I\kappa\tau_1} &= 
(-1)^{\tfrac12 (1-{\rm sgn}(\kappa))(w-1)}\, \varGamma(2-w)\, \left( q^{-\kappa} - \bar q^{\kappa} \, 
\sum_{\ell=0}^{-w} \frac{(4\pi\kappa\tau_2)^\ell}{\ell !} \right)
\\
&= (-1)^{\tfrac12 (1-{\rm sgn}(\kappa))(w-1)}\,\left[ \varGamma (2-w) - (1-w)\, \varGamma (1-w;4\pi\kappa \tau_2 ) \right] q^{-\kappa}\,, 
\end{split}
\end{equation}
and the remaining Fourier coefficients reduce to
\begin{equation}
\label{FourierFBOh}
\begin{split}
\tilde\cF_{m>0}(1-\tfrac{w}{2},\kappa,w)&= 4\pi\, \I^{-w} \, \varGamma(2-w)\, |\kappa|^{1-\tfrac{w}{2}}\, m^{\tfrac{w}{2}}\, \cZ_{1-\frac{w}{2}}(m,-\kappa) \, e^{-2\pi m\tau_2}\,,
\\
 \tilde\cF_{m<0}(1-\tfrac{w}{2},\kappa,w)&=4\pi\, \I^{-w}\, (1-w)\, |\kappa|^{1-\tfrac{w}{2}} \, 
 |m|^{\frac{w}{2}}  \cZ_{1-\frac{w}{2}}(m,-\kappa)\, \varGamma(1-w,-4\pi m\tau_2)\, e^{-2\pi m\tau_2}\,,
\\
 \tilde\cF_{m=0}(1-\tfrac{w}{2},\kappa,w)&=\frac{4\pi^2}{(2\pi\I)^w} \,\frac{\sigma_{1-w}(\kappa)}{\zeta(2-w)} \,.
\end{split}
\end{equation}
One thus recognises an expansion of the form \eqref{genharm} with coefficients
\begin{equation}
\begin{split}
& a_{-\kappa} = \varGamma(2-w)  \,,
\\
& a_{-\kappa<m<0} =0\,,
\\
& a_0 = \frac{4\pi^2}{(2\pi\I)^w} \,\frac{\sigma_{1-w}(\kappa)}{\zeta(2-w)}\,,
\\
& a_{m>0}=  4\pi\, \I^{-w} \, \varGamma(2-w)\, |\kappa|^{1-\frac{w}{2}}\, 
m^{\frac{w}{2}}\, \cZ_{1-\tfrac{w}{2}}(m,-\kappa) \,,
\\
& b_0 = 0\,,
\\
& b_{m>0} =
(1-w)\, |\kappa|^{1-w}\, \delta_{m,\kappa} + 
4\pi\, \I^{w}\, (1-w)\, |  m\,\kappa |^{1-\tfrac{w}{2}}\,   \cZ_{1-\tfrac{w}{2}}(m,\kappa)\,.
\end{split}
\end{equation}
In particular, $b_0=0$, so that the shadow of $\cF(1-\tfrac{w}{2},\kappa,w)$ is a cusp form of weight $2-w$, proportional to the holomorphic Poincar\'e series  $P(-\kappa,2-w)$.  Indeed, using \eqref{DwF} we find
\begin{equation}
\begin{split}
\bar D_w\cdot \cF(1-\tfrac{w}{2},\kappa,w) = &
\frac{1-w}{8\kappa}\, \cF(1-\tfrac{w}{2},\kappa,w-2)
\\
= & \frac{1-w}{8\kappa}\, (4\pi \, \kappa\, \tau_2)^{2-w}\, \overline{P(-\kappa,2-w)} \,,
\end{split}
\end{equation}
where in the second line we have recognised the Fourier expansion of the standard holomorphic Poincar\'e series of weight greater than 2. Similarly, using \eqref{DwF} the ghost of $\cF(1-\tfrac{w}{2},\kappa,w)$ is 
\begin{equation}
D_w^{1-w}\cdot \cF(1-\tfrac{w}{2},\kappa,w) = (2\kappa)^{1-w}\, \varGamma(2-w)\, 
\cF(1-\tfrac{w}{2},\kappa,2-w)\,,
\end{equation}
and corresponds to the Niebur-Poincar\'e series $\cF(\frac{w'}{2},\kappa,w')$, with $w'=2-w >2$  {\em within the  convergence domain}. Moreover, the Fourier expansion of the latter reproduces that of the Poincar\'e series $P(\kappa,w)$ of positive weight
\begin{equation}
\cF(\tfrac{w'}{2},\kappa,w') = q^{-\kappa} + 
2\pi\, \I^{-w'}\, \sum_{m>0} 
\, \left(  \frac{m}{\kappa}\right)^{\frac{w'-1}{2}} 
\sum_{c>0} \frac{S (m,-\kappa;c)}{c}\, I_{w'-1} \left( \frac{4\pi}{c} \sqrt{\kappa m}\right)\,q^m\ .
\end{equation}
As an aside, we note that the holomorphic part $\cF_a(1-\tfrac{w}{2},\kappa,w)$  of the Niebur-Poincar\'e series $\cF(s,\kappa,w)$ at $s=1-\tfrac{w}{2}$  reproduces the  Fourier expansion of the Poincar\'e series   $\varGamma(2-w)\, P(\kappa,w)$ defined by holomorphic regularisation as in \eqref{PoincaHol} and worked out in \cite{0306.30023,0695.10021}. Therefore, the non-holomorphic part $\cF_b(1-\tfrac{w}{2},\kappa,w)$ of the same  Niebur-Poincar\'e series provides the modular completion of the Eichler integral  $P(\kappa,w)$ --- a clear advantage of modular-invariant regularisation  over holomorphic regularisation.

To make this discussion less abstract, we shall now exhibit the harmonic Maass form $\cF(1-\tfrac{w}{2},\kappa,w)$, its shadow and its ghost for the two cases $w=-10$ and  $w=-14$ (lower values of $|w|$ will be discussed in the next subsection) and $\kappa=1$.  Evaluating the Fourier coefficients numerically, we find:
\begin{itemize}
\item For $w=-10$, 
\begin{equation}
\cF(6,1,-10) = \cF_{b}(6,1,-10)+ 11! \left[ q^{-1} - \frac{65520}{691} -1842.89\, q -23274.08 \,q^2+\dots \right]
\label{ex1}
\end{equation}
where $\cF_{b}(6,1,-10)$ is the non-holomorphic component. The  shadow of \eqref{ex1} reads
\begin{equation}
\cF(6,1,-12) = (4\pi \tau_2)^{12}\, \overline{P(-1,12)}\ ,\qquad
P(-1,12) = \beta_{12}\, \varDelta\ ,
\end{equation}
where the modular discriminant $\varDelta$ generates the space of cusp forms of weight 12. The ghost, obtained by acting with $D^{11}$ on the holomorphic part, can be written as 
\begin{equation}
\begin{split}
\cF(6,1,12) &=\frac{83 E_4^3 E_6^2-11 E_6^4}{72 \varDelta}+\alpha_{12}\, \frac{(E_4^3-E_6^2)^2}{\varDelta}
\\
&= q^{-1} + 1842.89\, q + 47665306.53\, q^2 +\dots \,,
\end{split}
\end{equation}
where $\alpha_{12}=0.201029508104\dots, \beta_{12}=2.840287517\dots$ are irrational numbers. The coefficients $1842.89$, $23274.08$, $47665306.53$ are two-digit approximations of the exact values $324(9216\, \alpha_{12}-1847)$, $(60617 -69984\, \alpha_{12})/2$, $1024(60617 -69984\, \alpha_{12})$, respectively. This example was discussed in detail in \cite{OnoMockDelta}.

\item Similarly, for $w=-14$,
\begin{equation}
\cF(8,1,-14) = \cF_b(8,1,-14)+ 15! \left[ q^{-1} -\frac{16320}{3617}- 45.67 \,q - 366.47\, q^2 +\dots\right]
\label{pippo}
\end{equation}
where $\cF_{b}(8,1,-14)$ is the non-holomorphic component. The shadow of \eqref{pippo} reads
\begin{equation}
\cF(8,1,-16) = (4\pi \tau_2)^{16}\, \overline{P(-1,16)}\ ,\qquad
P(-1,16)=\beta_{16} \, E_4\, \varDelta
\end{equation}
where $E_4 \varDelta$ generate the space of cusp forms of weight 16. The ghost, obtained by acting with $D^{15}$ on the holomorphic part, can be written as 
\begin{equation}
\begin{split}
\cF(8,1,16) &= \frac{73 E_4^4 E_6^2-E_4 E_6^4}{72\varDelta}+\alpha_{16}\, 
\frac{E_4^7-2 E_4^4 E_6^2+E_4 E_6^4}{\varDelta} 
\\
&= q^{-1}+45.67\, q+12008361.57\, q^2+\dots \,,
\end{split}
\end{equation}
where $\alpha_{16}=0.137975847804\dots$ and $\beta_{16}=1.3061364711\dots$ are irrational numbers. The coefficients  $45.67$, $366.47$ in Eq. \eqref{pippo} are two-digit approximations of the exact values $36 ( 82944\, \alpha_{16} - 11443)$, $(314928\,\alpha_{16}-37589)/16$, respectively.
\end{itemize}

These two examples illustrate the fact that Fourier coefficients of harmonic Maass forms are in general irrational numbers.

\subsection{Weak holomorphic modular forms from Niebur-Poincar\'e series} 

We  now come to our main goal, {\em i.e.} to find an absolutely convergent Poincar\'e series  representation of any weak holomorphic modular form $\varPhi_w$ of weight $w\leq 0$ and $\kappa$-order pole at the cusp, with given principal part
\begin{equation}
\label{defPhi}
\varPhi^-_w(\tau)=\sum_{-\kappa\leq m <0} a_m\, q^m \,.
\end{equation}
As we shall see, any such $\varPhi_w$ can be expressed as a linear combination of the Niebur-Poincar\'e series $\cF(s,\kappa,w)$. 

We have noted  in the previous subsection that the eigenvalue  of a weak holomorphic modular form under the hyperbolic Laplacian $\Delta_w$ coincides with the eigenvalue of the Niebur-Poincar\'e series  whenever $s=1-\tfrac{w}{2}$. At this value, however, $\cF(1-\tfrac{w}{2},\kappa,w)$ is a weak harmonic Maass form, in general not holomorphic. Exceptions to this statement occur at the special values $w\in \{-2,-4,-6,-8, -12\}$, where the space of holomorphic cusp forms of weight $2-w$ is empty,
and ${\mathcal F} (1-\tfrac{w}{2}, \kappa , w)$ can be recognised as an element of the ring of weak holomorphic modular forms by matching the principal part of their expansions.
For $\kappa =1$ the exact identification is reported in Table \ref{holtable}, while for $\kappa>1$, the proper identification of ${\mathcal F}(1-\frac{w}{2} ,\kappa,w)$ can be obtained by  acting on ${\mathcal F}(1-\frac{w}{2} ,1,w)$ with the Hecke operator $T_\kappa$, as given by Eq. \eqref{HeckeF1}.

\begin{table}
\centering 
\begin{tabular}{c | c c }
\tcolrow   & & 
\\
\tcolrow $w$ & ${\mathcal F} (1-\tfrac{w}{2}, 1,w)$ & ${\mathcal F} (1-\tfrac{w}{2}, 1,2-w)$
\\[4mm]
\hline 
\tcolrow & & 
\\
\tcolrow $0$ & $j+24$ & $ E_4^2 E_6\, \varDelta^{-1} $
\\[2mm]
\tcolrow $-2$ &  $3!\, E_4 E_6\, \varDelta^{-1}$ &  $E_4 (j-240) $
\\[2mm]
\tcolrow $-4$ &   $5!\, E_4^2 \, \varDelta^{-1}$ & $E_6 (j+204)$
\\[2mm]
\tcolrow $-6$ &  $7!\, E_6 \,\varDelta^{-1}$ & $E_4^2 (j-480) $
\\[2mm]
\tcolrow $-8$ &  $9!\, E_4 \,\varDelta^{-1}$ & $E_4 E_6 (j+264)$
\\[2mm]
\tcolrow $-12$ &  $13!\, \varDelta^{-1}$ & $ E_4^2 E_6 (j+24)  $
\\
\tcolrow & & 
\end{tabular}
\caption{Weak holomorphic modular forms obtained as the limit $s\to 1-\frac{w}{2}$ of ${\mathcal F} (s,1,w)$, for the values $w\in\{0,-2,-4,-6,-8,-12\}$. For $w$ negative and outside this range, the limit yields a weak harmonic Maass form. The second line shows the ghost, which is a  weak holomorphic modular form of weight $2-w$ with vanishing constant term (aside from the case $w=0$)}    
\label{holtable}
\end{table}

For $w\leq 0$ outside the list above, the space of cusp forms of weight $2-w$ is not empty, and $\cF(1-\tfrac{w}{2},\kappa,w)$ is indeed a genuine harmonic Maass form, with non-vanishing shadow.  Nevertheless, it can be shown \cite{1004.11021} that the linear combination
\begin{equation}
\label{Flin}
 \cG(s,w)\equiv \frac{1}{\varGamma(2-w)}\, 
\sum_{-\kappa\leq m<0} \, a_m \,   \cF(s,m,w)\ ,\qquad 
\end{equation}
with coefficients $a_m$ determined by the principal part 
\begin{equation}
\varPhi_w^-=\sum_{-\kappa\leq m<0} \, a_m \, q^{-m}
\end{equation}
of any  weak holomorphic form $\varPhi_w$ of negative weight $w$, reduces to a weak holomorphic modular form for  $s=1-\tfrac{w}{2}$, namely  $\varPhi_w$ itself.  Said differently, the shadows of the weak harmonic Maass forms $\cF(1-\tfrac{w}{2},\kappa,w)$ cancel in the linear combination \eqref{Flin}.  As a result, any $\varPhi_w$ can be represented as the linear combination
\begin{equation}
\varPhi_w = \frac{1}{\varGamma(2-w)}\, 
\sum_{-\kappa\leq m<0} \, a_m \,   \cF(1-\tfrac{w}{2},m,w)\ ,
\label{bruinier}
\end{equation}
or equivalently, using \eqref{Mharm}, as an absolutely convergent Poincar\'e sum
\begin{equation}
\label{PhiModReg}
\varPhi_w=\tfrac12 \sum_{\gamma\in \Gamma_\infty\backslash \Gamma}\, 
\left(\varPhi_w^- - \sum_{-\kappa\leq m<0} \sum_{\ell=0}^{-w} \, a_m \, \bar q^{m} \, 
 \frac{(4\pi\kappa\tau_2)^\ell}{\ell !} \right)\, \Bigg|_w \, \gamma \,,
\end{equation}
where the subtraction in the bracket ensures that the seed is $\cO(\tau_2^{1-\frac{w}{2}})$ as $\tau_2\to 0$. We stress that,  unlike the holomorphic regularisation in \eqref{PoincaHol},
the expression \eqref{PhiModReg} is manifestly modular covariant and absolutely convergent.

To illustrate the power of Eq. \eqref{bruinier}, let us reconsider the two examples of  the previous subsection, now allowing for $\kappa=1,2$.

\begin{itemize}
\item 
For $w=-10$, $\cF(6,2,-10)$ and $\cF(6,1,-10)$ are separately weak harmonic Maass forms with irrational coefficients, but the linear combination
\begin{equation}
\cF(6,2,-10)+24\, \cF(6,1,-10) = 11! \, \tfrac{E_4^2 E_6}{\Delta^2} = 
11! \, ( q^{-2} + 24\, q^{-1} - 196560 + \dots )
\end{equation}
produces (up to an overall normalisation) the unique weak holomorphic form\footnote{Compare the simplicity of our expression to the corresponding equation in Sec 4.1 of \cite{OnoMockDelta}.} of weight $-10$ with a double pole at $q=0$;
\item 
Similarly, for  $w=-14$, $\cF(8,2,-14)$ and $\cF(8,1,-14)$ are separately 
weak harmonic Maass forms with irrational coefficients, but the linear combination
\begin{equation}
\cF(8,2,-14)-216\, \cF(8,1,-14) = 15! \, \tfrac{E_4 E_6}{\Delta^2} = 
15! \, ( q^{-2} - 216\, q^{-1} - 146880 + \dots )
\end{equation}
produces (up to an overall normalisation) the unique weak holomorphic form of weight $-14$ 
with a double pole at $q=0$.
\end{itemize}

Similar relations occur for higher negative weight $w<-14$ and higher order $\kappa$
of the pole at $q=0$. 

\subsection{Weak almost  holomorphic modular forms from Niebur-Poincar\'e series}

For physics applications it is important to extend our previous analysis to the case of weak {\it almost} holomorphic modular forms, {\em i.e.} elements of the ring generated by the almost holomor\-phic Eisenstein series $\hat E_2$ and the ordinary weak holomorphic modular forms, or equivalently, by the modular derivatives  $D^n \varPhi$ of ordinary weak holomorphic modular forms.

To this end, it is important to note that for any integer $n\geq 0$, it  follows from  \eqref{DwF} that the Niebur-Poincar\'e series $\cF(s,\kappa,w)$ evaluated at the point $s= 1-\tfrac{w}{2}+n$ can be expressed as
\begin{equation}
\cF(1-\tfrac{w}{2}+n,\kappa,w) = \frac{1}{(2\kappa)^n\, n!}\, D^{n}\, 
\cF(1-\tfrac{w}{2}+n, \kappa,w-2n)\ ,
\end{equation}
where $D^n$ is the iterated modular derivative \eqref{iteratedD}. The Niebur-Poincar\'e series $\cF(s',\kappa,w')$ appearing on the {\em r.h.s.} satisfies $s'=1-\tfrac{w'}{2}$, and thus is a  harmonic Maass  form. 

As a result, provided that the coefficients $a_m$ in the linear combination \eqref{Flin} are chosen such that 
\begin{equation}
\varPhi_{w-2n}^-\equiv \sum_{-\kappa\leq m<0} \,  \frac{a_m}{(2m)^n\, n!} \, q^{m} 
\end{equation}
is the principal part of a weak holomorphic modular form $\varPhi_{w-2n}$ of weight $w-2n$, then  
the linear combination ${\mathcal G} (s,w)$ in \eqref{Flin} evaluated  at the point  $s=1-\tfrac{w}{2}+n$ reproduces an almost holomorphic modular form of weight $w$,
\begin{equation}
{\mathcal G} (1-\tfrac{w}{2}+n,w)=\frac{1}{\varGamma (2-w)}\, \sum_{-\kappa\leq m<0}  a_m\, \cF(1-\tfrac{w}{2}+n,m,w) = D^n\, \varPhi_{w-2n} \ .
\end{equation}
More generally, we refer to the space $\bigoplus_{n\geq 0} \cH(1-\tfrac{w}{2}+n ,w)$ as the space of ``weak almost  harmonic Maass forms'', of which almost holomorphic modular forms are only a subspace. The general Fourier expansion of such forms can be obtained by taking the limit $s=1-\tfrac{w}{2}+n$ in eqs. \eqref{FskwF} and \eqref{FourierFBO}, and by using the identities \eqref{Mn} and \eqref{Wn}.
Similarly, the series $\cF(s,\kappa ,w)$ at the point $s=1-\tfrac{w}{2}+n$ for $n\leq -2$ may  be obtained from  $\cF(s', \kappa , w')$ at the point $s'=-\tfrac{w'}{2}$ by using the lowering operator $\bar D_w$.

\begin{figure}
\begin{center}
\begin{picture}(320,260)(-160,-10)
\linethickness{0.3mm}
\put(-150,0){\line(1,0){300}}\put(150,0){\vector(1,0){0.2}}
\linethickness{0.3mm}
\put(0,-10){\line(0,1){270}}\put(0,260){\vector(0,1){0.2}}
\linethickness{0.2mm}
\multiput(-5,-5)(1.4,1.4){100}{\multiput(0,0)(0.12,0.12){6}{\line(1,0){0.12}}}
\linethickness{0.2mm}
\multiput(5,-5)(-1.4,1.4){100}{\multiput(0,0)(-0.12,0.12){6}{\line(1,0){0.12}}}
\linethickness{0.2mm}
\multiput(-65,-5)(1.4,1.4){150}{\multiput(0,0)(0.12,0.12){6}{\line(1,0){0.12}}}
\linethickness{0.2mm}
\multiput(65,-5)(-1.4,1.4){150}{\multiput(0,0)(-0.12,0.12){6}{\line(1,0){0.12}}}
\linethickness{0.2mm}
\multiput(-140,60)(0.5,0){560}{\line(1,0){0.1}}
\linethickness{0.4mm}
\put(15,120){\line(1,0){30}}\put(45,120){\vector(1,0){0.2}}
\put(-15,120){\line(-1,0){30}}\put(-45,120){\vector(-1,0){0.2}}
\put(0,0){\makebox(0,0)[c]{$\bullet$}}
\put(0,60){\makebox(0,0)[c]{$\bullet$}}
\put(0,120){\makebox(0,0)[c]{$\bullet$}}
\put(0,180){\makebox(0,0)[c]{$\bullet$}}
\put(0,240){\makebox(0,0)[c]{$\bullet$}}
\put(60,0){\makebox(0,0)[c]{$\bullet$}}
\put(60,60){\makebox(0,0)[c]{$\bullet$}}
\put(60,120){\makebox(0,0)[c]{$\bullet$}}
\put(60,180){\makebox(0,0)[c]{$\bullet$}}
\put(60,240){\makebox(0,0)[c]{$\bullet$}}
\put(120,0){\makebox(0,0)[c]{$\bullet$}}
\put(120,60){\makebox(0,0)[c]{$\bullet$}}
\put(120,120){\makebox(0,0)[c]{$\bullet$}}
\put(120,180){\makebox(0,0)[c]{$\bullet$}}
\put(120,240){\makebox(0,0)[c]{$\bullet$}}
\put(-60,0){\makebox(0,0)[c]{$\bullet$}}
\put(-60,60){\makebox(0,0)[c]{$\bullet$}}
\put(-60,120){\makebox(0,0)[c]{$\bullet$}}
\put(-60,180){\makebox(0,0)[c]{$\bullet$}}
\put(-60,240){\makebox(0,0)[c]{$\bullet$}}
\put(-120,0){\makebox(0,0)[c]{$\bullet$}}
\put(-120,60){\makebox(0,0)[c]{$\bullet$}}
\put(-120,120){\makebox(0,0)[c]{$\bullet$}}
\put(-120,180){\makebox(0,0)[c]{$\bullet$}}
\put(-120,240){\makebox(0,0)[c]{$\bullet$}}
\put(160,0){\makebox(0,0)[c]{$w$}}
\put(0,270){\makebox(0,0)[c]{$s$}}
\put(60,-10){\makebox(0,0)[c]{$2$}}
\put(120,-10){\makebox(0,0)[c]{$4$}}
\put(-60,-10){\makebox(0,0)[c]{$-2$}}
\put(-120,-10){\makebox(0,0)[c]{$-4$}}
\put(-10,60){\makebox(0,0)[c]{$1$}}
\put(-10,120){\makebox(0,0)[c]{$2$}}
\put(-10,180){\makebox(0,0)[c]{$3$}}
\put(30,130){\makebox(0,0)[c]{$D$}}
\put(-30,130){\makebox(0,0)[c]{$\bar{D}$}}
\put(0,45){\makebox(0,0)[c]{$j+24$}}
\put(60,45){\makebox(0,0)[c]{$E_4^2\,E_6/\varDelta$}}
\put(0,210){\makebox(0,0)[c]{weak almost  harmonic}}
\put(120,135){\rotatebox{45.0}{\makebox(0,0)[c]{$s=\tfrac{w}{2}$: weak hol. (ghost)}}}
\put(100,195){\rotatebox{45.0}{\makebox(0,0)[c]{weak almost  hol.}}}
\put(-120,135){\rotatebox{-45.0}{\makebox(0,0)[c]{$s=-\tfrac{w}{2}$: $\tau_2^{2-w}\times\,$anti-hol. (shadow)}}}
\put(-120,195){\rotatebox{-45.0}{\makebox(0,0)[c]{$s=1-\tfrac{w}{2}$: weak harmonic}}}
\end{picture}
\end{center}
\caption{Phase diagram for the Niebur-Poincar\'e series $\cF(s,\kappa,w)$ for integer
values of $(\frac{w}{2},s)$ with $s\geq 1$. For low negative values of $w$, $\cF(s,\kappa,w)$
reduces to an ordinary weak almost  holomorphic Maass form, see Table \ref{tabcone}.
\label{lightconefig}}
\end{figure}
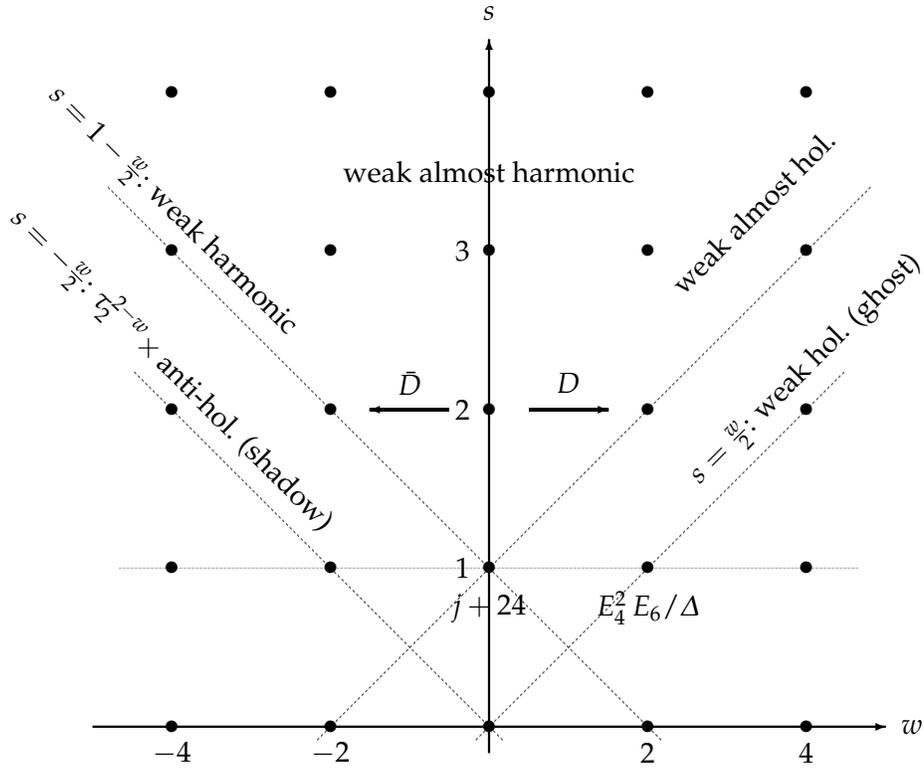

\subsection{Summary}

To summarise this discussion, it is useful to consider the  plane of the variables $(\tfrac{w}{2},s)$
as in Figure \ref{lightconefig}.
The Niebur-Poincar\'e series $\cF(s,\kappa,w)$ 
converges absolutely for $s>1$. For integer values of $s$, it is
generally a weak almost  harmonic Maass form, and on the line $s=1-\frac{w}{2}$ (and $w<0$), 
$\cF(s,\kappa,w)$ becomes a weak harmonic Maass form. On the line $s=-\frac{w}{2}$,
obtained from the former by acting with the lowering operator $\overline{D}$, 
$\cF(s,\kappa,w)$ reduces, up to an overall multiplicative factor $\tau_2^{-w}$, to the complex conjugate of a cusp  form of weight $2-w$, known as the shadow of the harmonic Maass form $\cF(s,\kappa,w+2)$.
On the line  $s=\frac{w}{2}$, $\cF(s,\kappa,w)$ is instead a weak holomorphic modular form. It is connected to its expression on the line $s=1-\frac{w}{2}$ by the  action of the iterated
raising operator $D^{1-w}$, and thus we refer to it as the `ghost' of the harmonic Maass form $\cF(s,\kappa,2-w)$. In the 
quadrant $w>2, s>1$, $\cF(s,\kappa,w)$ is more generally a weak almost  holomorphic
modular form. For low negative values of $w$ and $s$ integer, $\cF(s,\kappa,w)$ 
is in fact always a weak almost  holomorphic
modular form,  as displayed in Table
\ref{tabcone}. Genuine harmonic Maass forms start appearing at $s=6$ and $s\geq 8$.

\begin{sidewaystable}
\centering

\begin{tabular}{c|ccccccccccc}
\\
$s\backslash w$ 
& $-10$ 
& $ -8$ 
& $-6$ 
& $-4$ 
& $-2$ 
& $0$ 
& $2$ 
& $4$ 
& $6$ 
& $8$ 
& $10$ 
\\ [4mm]
\hline
\\ 
$5$ 
& $0$ 
&  $9!  \tfrac{E_4}{\Delta}$
& $\tfrac{9!}{2}  D \tfrac{E_4}{\Delta}$ 
& $\tfrac{9!}{8} \, D^2 \tfrac{E_4}{\Delta}$ 
&  $\tfrac{9!}{2^3\, 3!} D^3 \tfrac{E_4}{\Delta}$ 
& $\tfrac{9!}{2^4\, 4!}  D^4 \tfrac{E_4}{\Delta}$ 
& $\tfrac{9!}{2^5\, 5!}  D^5 \tfrac{E_4}{\Delta} $
& $\tfrac{9!}{2^6\, 6!}  D^6 \tfrac{E_4}{\Delta} $
& $\tfrac{9!}{2^7\, 7!}  D^7 \tfrac{E_4}{\Delta} $
& $\tfrac{9!}{2^7\, 8!}  D^8 \tfrac{E_4}{\Delta} $
& $E_4 E_6 (j+264) $
\\[3mm]
$4$ 
& $0$ 
& $0$ 
& $7!  \tfrac{E_6}{\Delta}$
& $\tfrac{7!}{2}  D \tfrac{E_6}{\Delta} $
& $\tfrac{7!}{8} \, D^2 \tfrac{E_6}{\Delta} $
& $\tfrac{7!}{2^3\, 3!} D^3 \tfrac{E_6}{\Delta} $
& $\tfrac{7!}{2^4\, 4!}  D^4 \tfrac{E_6}{\Delta} $
& $\tfrac{7!}{2^5\, 5!}  D^5 \tfrac{E_6}{\Delta} $
& $\tfrac{7!}{2^6\, 6!}  D^6 \tfrac{E_6}{\Delta} $
& $E_4^2 (j-480) $
& $\tfrac{7!}{2^8\, 8!}  D^8 \tfrac{E_6}{\Delta} $
\\[3mm]
$3$ 
& $0$
& $0$ 
& $0$ 
& $5! \tfrac{E_4^2}{\Delta} $
& $\tfrac{5!}{2} D\tfrac{E_4^2}{\Delta} $
& $\tfrac{5!}{8}  D^2 \tfrac{E_4^2}{\Delta} $
& $\tfrac{5!}{2^3 3!} D^3 \tfrac{E_4^2}{\Delta} $
& $\tfrac{5!}{2^4 4!} D^4 \tfrac{ E_4^2}{\Delta} $
& $E_6 (j+504)  $
& $\tfrac{5!}{2^6 6!}  D^6 \tfrac{E_4^2}{\Delta} $
& $\tfrac{5!}{2^7 7!}  D^7 \tfrac{E_4^2}{\Delta} $
\\[3mm]
$2$ 
& $0$ 
& $0$ 
& $0$ 
& $0$ 
& $3!\, \tfrac{E_4 E_6}{\Delta} $
& $3 D \tfrac{E_4 E_6}{\Delta} $
& $\tfrac{3}{4} D^2 \tfrac{E_4 E_6}{\Delta} $
& $E_4 (j-240) $
& $\tfrac{3!}{2^4 4!} D^4\tfrac{E_4 E_6}{\Delta} $
& $\tfrac{3!}{2^5 5!} D^5 \tfrac{E_4 E_6}{\Delta} $
& $\tfrac{3!}{2^6 6!} D^6 \tfrac{E_4 E_6}{\Delta} $
\\[3mm]
$1$ 
& $0$ 
& $0$ 
& $0$ 
& $0$ 
& $0$ 
& $j+24$ 
& $\tfrac{E_4^2 E_6}{\Delta} $
& $\tfrac{1}{2^2 2!} D^2 j $
& $\tfrac{1}{2^3 3!} D^3 j $
& $\tfrac{1}{2^4 4!} D^4 j $
& $\tfrac{1}{2^5 5!} D^5 j $
\\
\\
\end{tabular}
\caption{Niebur-Poincar\'e series $\cF(s,1,w)$ at the special values
$s=1-\tfrac{w}{2}+n$ with $n$ integer, for low negative values of $w$. \label{tabcone}}
\end{sidewaystable}

\section{A new road to one-loop modular integrals}
\label{integrals}

We are interested in the evaluation of one-loop modular integrals of the form \eqref{modintegral}, while keeping manifest at all steps the automorphisms of the Narain lattice, {\em i.e.}  T-duality. Such integrals encode, for instance, threshold corrections to the running of gauge and gravitational couplings. The function $\varPhi$, related to the elliptic genus and dependent on the vacuum under consideration, is in general a weak almost  holomorphic modular form of non-positive weight. For example, 
in ${\mathcal N}=4$ compactifications of the ${\rm SO} (32)$ heterotic string 
(with vanishing Wilson lines) one finds a linear combination of zero-weight
weak almost  holomorphic modular forms
\cite{Lerche:1987qk}
\begin{equation}
\begin{split}
\varPhi (\tau ) &=t_8 \, {\rm tr} F^4 + \frac{1}{2^7\, 3^2\, 5}\, \frac{E_4^3}{\varDelta} \, t_8 \, {\rm tr} R^4 + \frac{1}{2^9\, 3^2} \frac{\hat E_2^2\, E_4^2}{\varDelta} \, t_8 ({\rm tr} R^2 )^2 
\\
& + \frac{1}{2^8\, 3^2} 
\left(  \frac{\hat E_2 \, E_4 \, E_6}{\varDelta} - \frac{\hat E_2^2 \, E_4^2}{\varDelta} \right) \, t_8 \, {\rm tr} F^2\, {\rm tr} \, R^2
\\
& + \frac{1}{2^9\, 3^2}  \left( \frac{E_4^3}{\varDelta} + \frac{\hat E_2^2\, E_4^2}{\varDelta} -2\, \frac{\hat E_2 \, E_4 \, E_6}{\varDelta} - 2^7\, 3^2 \right) \, t_8\, ( {\rm tr} \, F^2)^2 \,,
\end{split}
\label{N4thresholds}
\end{equation}
where $t_8$ is the familiar tensor appearing in four-point amplitudes of the heterotic string, and $F$ and $R$ are the gauge field strength and curvature two-form. A similar expression arises
for gauge and gravitational couplings in the ${\rm E}_8\times {\rm E}_8$ heterotic string.

While the traditional procedure for evaluating integrals of the form \eqref{modintegral} has been to unfold the integration domain $\cF$ against the lattice partition function $\varGamma_{d+k,d}$, 
 in \cite{Angelantonj:2011br} we instead proposed  to represent $\varPhi$ 
as a Poincar\'e series of the form \eqref{Pseed}, which is then amenable
to the unfolding procedure. 
The advantage of this approach is that T-duality is kept manifest at all steps and the final result is expressed as a sum over BPS states which is manifestly invariant under ${\rm O} (d+k,d;\mathbb{Z})$. Moreover, singularities associated to states becoming massless 
at special points in the Narain moduli space are easily read off from this representation. 

\subsection{Niebur-Poincar\'e series in a nutshell}

In order to implement this strategy, it is essential to represent
$\varPhi$ as an \emph{absolutely convergent}  Poincar\'e series, so 
that the unfolding of the fundamental domain is justified. 
Fortunately, as discussed in detail in Section \ref{sec_Niebur} and summarised in the following, any weak almost  holomorphic
modular form $\varPhi_{w}$ of weight $w\leq 0$  can be written as a linear combination 
of  Niebur-Poincar\'e series, defined as 
\begin{equation}
\label{defNiebur2}
\begin{split}
{\mathcal F} (s,\kappa,w) =&\tfrac12 \sum_{\gamma\in \Gamma_\infty\backslash \Gamma} \,
\cM_{s,w}(-\kappa\tau_2)\, e^{-2\pi\I\kappa\tau_1}\, \vert_w\, \gamma \\
=& \tfrac{1}{2} \sum_{(c,d)=1} (c\tau + d)^{-w} \, {\mathcal M}_{s,w} \left( \frac{-\kappa\tau_2}{|c\tau + d|^2} \right) \, \exp \left\{ - 2 \I \pi \kappa \left( \frac{a}{c} - \frac{c\tau_1 + d}{c |c\tau +d|^2} \right)\right\}\, .
\end{split}
\end{equation}
Here $\cM_{s,w}$ is related to the Whittaker $M$-function via
\begin{equation}
{\mathcal M}_{s,w} (-y ) = (4\pi y )^{-w/2}\, M_{-\frac{w}{2} , s-\frac{1}{2}} (4\pi y )\,,
\end{equation}
and $s$ is a complex parameter, the real part of which must be larger than 1 for absolute convergence. The choice of the Whittaker function in \eqref{defNiebur2} is dictated by the requirement that $\cF(s,\kappa,w)$ be an eigenmode of the hyperbolic Laplacian $\Delta_w$ (see Eq. \eqref{laplFskw}), and 
behave as $q^{-\kappa}$  at the cusp $q\equiv e^{2\pi\I\tau}= 0$ (see Eq. \eqref{limMcusp}),
thus reproducing, for $\kappa=1$, the simple pole 
associated to the unphysical tachyon of the heterotic string. The set of Niebur-Poincar\'e series ${\mathcal F} (s,\kappa,w)$ is closed under
the action of the derivative operators $D_w$ and $\bar D_w$ defined in
\eqref{modularderiv}, which, according to \eqref{DwF}, act by raising or lowering the weight $w$ by two units while keeping $s$ fixed.

At the special point $s=1-\tfrac{w}{2}$, which for $w<0$ lies within the  domain of absolute convergence, 
the Niebur-Poincar\'e series ${\mathcal F} (s,\kappa,w)$ becomes a weak harmonic 
Maass form\footnote{For a definition of weak harmonic 
Maass forms see Section \ref{defMaass}.}.
In particular, unless $w$ takes one 
of the special values listed in Table \ref{holtable}, it is in general not holomorphic. Although the values listed in the table essentially exhaust all the cases of interest in string theory, it is a remarkable fact 
that  linear combinations of Niebur-Poincar\'e series, 
with coefficients determined by the principal part of a weak holomorphic modular form $\varPhi_w$,
are in fact weakly holomorphic, and reproduce $\varPhi_w$ itself \cite{1004.11021}:
\begin{equation}
\label{Phiexp}
\varPhi_w^-=\sum_{-\kappa\leq m<0} \, a_m \, q^{-m}\quad 
\Rightarrow \quad 
\varPhi_w = \frac{1}{\varGamma(2-w)}\, 
\sum_{-\kappa\leq m<0} \, a_m \,   \cF(1-\tfrac{w}{2},m,w) \ .
\end{equation}
Moreover, upon using \eqref{DwF} one can also relate weak
almost  holomorphic modular forms involving (up to) $n$ powers
of $\hat E_2$ --- or equivalently, obtained by acting up to $n$ times  with the derivative
operator $D_w$ on  a weak holomorphic modular form --- to linear 
combinations of ${\mathcal F} (s,\kappa,w)$ evaluated at the special points $s=1-\tfrac{w}{2}+n'$, 
with $0\leq n'\leq n$.

In the cases relevant to heterotic string threshold corrections, the elliptic genus $\varPhi_w$
has a simple pole at $q=0$, corresponding to the unphysical tachyon, and therefore the expansion
\eqref{Phiexp} includes only one term, with $\kappa=m=1$ (modulo an additive
constant in the case $w=0$). Moreover, the weight $w$
is related to the signature $(d+k,d)$ of the Narain lattice by $w=-k/2$. Since string theory restricts the Narain lattice to be even and self-dual, so that $\varGamma_{d+k,d}$ is covariant
under the full modular group $\varGamma={\rm SL}(2,\IZ)$, the possible values of $w$ 
are $w=0$ (corresponding to the point of unbroken ${\rm E}_8 \times {\rm E_8}$ or ${\rm SO}(32)$
gauge symmetry), $w=-4$ (corresponding to the point of unbroken ${\rm E}_8$ symmetry,
with arbitrary Wilson lines for the other ${\rm E}_8$ factor), or $w=-8$ (corresponding to generic
values of the Wilson lines in ${\rm E}_8 \times {\rm E_8}$ or ${\rm SO}(32)$). The complete
list of weak almost  holomorphic modular forms with a simple pole at $q=0$ and  modular
weights $w=0,-2,-4,-6,-8,-10$,  together with their expressions as linear combinations
of Niebur-Poincar\'e series, can be found in Table \ref{maintable}.
Although string-theory applications only require $\kappa=1$, our methods apply
equally well for arbitrary positive integer values of $\kappa$, which we 
therefore keep general until Section \ref{sec_sing}.

\begin{table}
\centering
\begin{tabular}{c }
\tcolrow $w=0$
\\[2mm]
\hline 
\\[-2mm]
$
\begin{array}{r l}
\frac{\hat E_2 E_4 E_6}{\varDelta} =& {\mathcal F} (2,1,0) - 5 \, {\mathcal F} (1,1,0) -144
\\[1.5mm]
\frac{\hat E_2^2 E_4^2}{\varDelta} =& \frac{1}{5}\, {\mathcal F} (3,1,0) -4{\mathcal F} (2,1,0) +13 \, {\mathcal F} (1,1,0) +144
\\[1.5mm] 
\frac{\hat E_2^3 E_6}{\varDelta} =& \frac{3}{175}\, {\mathcal F} (4,1,0) - 
\frac{3}{5}\, {\mathcal F} (3,1,0) +\frac{33}{5}\, {\mathcal F} (2,1,0) -17 \, {\mathcal F} (1,1,0) -144
\\[1.5mm]
\frac{\hat E_2^4 \, E_4}{\varDelta} =& \frac{1}{1225} \, {\mathcal F} (5,1,0) - \frac{6}{175} \, {\mathcal F} (4,1,0) + \frac{18}{35} \, {\mathcal F} (3,1,0) - \frac{16}{5}\, {\mathcal F} (2,1,0) 
\\[1.5mm]
&+ \frac{29}{5}\, {\mathcal F} (1,1,0) +\frac{144}{5}
\\[1.5mm]
\frac{\hat E^6_2}{\varDelta} =& \frac{1}{1926925} {\mathcal F} (7,1,0) - \frac{3}{2695}{\mathcal F} (5,1,0) + \frac{6}{175} {\mathcal F} (4,1,0) - \frac{3}{7} {\mathcal F} (3,1,0) 
\\[1.5mm]
&+ \frac{12}{5} {\mathcal F} (2,1,0) 
- \frac{29}{7} {\mathcal F} (1,1,0) - \frac{144}{7}
\end{array}
$
\\[24mm]
\tcolrow $w=-2$
\\[2mm]
\hline
\\[-2mm]
$
\begin{array}{r l}
\frac{\hat E_2 E_4^2}{\varDelta} =& \frac{1}{40} {\mathcal F} (3,1,-2) - \frac{1}{3} {\mathcal F} (2,1,-2) 
\\[1.5mm]
\frac{\hat E_2 ^2 E_6}{\varDelta} =& \frac{1}{525} {\mathcal F} (4,1,-2) - \frac{1}{20} {\mathcal F} (3,1,-2) + \frac{11}{30} {\mathcal F} (2,1,-2)
\\[1.5mm]
\frac{\hat E_2^3 E_4}{\varDelta} =& \frac{1}{11760} {\mathcal F} (5,1,-2) - \frac{1}{350} {\mathcal F} (4,1,-2) + \frac{9}{280} {\mathcal F} (3,1,-2) - \frac{2}{15} {\mathcal F} (2,1,-2)
\\[1.5mm]
\frac{\hat E_2^5}{\varDelta} =& \frac{1}{19819800} {\mathcal F} (7,1,-2) - \frac{1}{12936} {\mathcal F} (5,1,-2) +\frac{1}{525} {\mathcal F} (4,1,-2) - \frac{1}{56} {\mathcal F} (3,1,-2)
\\[1.5mm]
&+ \frac{1}{15} {\mathcal F} (2,1,-2)
\end{array}
$
\\[17mm]
\tcolrow $w=-4$
\\[2mm]
\hline
\\[-2mm]
$
\begin{array}{r l}
\frac{\hat E_2 E_6}{\varDelta} =& \frac{1}{2520} {\mathcal F} (4,1,-4) - \frac{1}{120} {\mathcal F} (3,1,-4)
\\[1.5mm]
\frac{\hat E_2^2 E_4}{\varDelta} =& \frac{1}{70560} {\mathcal F} (5,1,-4) - \frac{1}{2520} {\mathcal F} (4,1,-4) +\frac{1}{280} {\mathcal F} (3,1,-4 )
\\[1.5mm]
\frac{\hat E_2^4}{\varDelta} =& \frac{1}{148648500} {\mathcal F} (7,1,-4) - \frac{1}{129360} {\mathcal F} (5,1,-4) + \frac{1}{6300} {\mathcal F} (4,1,-4 ) - \frac{1}{840} {\mathcal F} (3,1,-4 )
\end{array}
$
\\[11mm]
\tcolrow $w=-6$
\\[2mm]
\hline
\\[-2mm]
$
\begin{array}{r l}
\frac{\hat E_2 E_4}{\varDelta} =& \frac{1}{241920} {\mathcal F} (5,1,-6) - \frac{1}{10080} {\mathcal F} (4,1,-6)
\\[1.5mm]
\frac{\hat E_2^3}{\varDelta} = & \frac{1}{792792000} {\mathcal F} (7,1,-6) - \frac{1}{887040} {\mathcal F} (5,1,-6) +\frac{1}{50400} {\mathcal F} (4,1,-6 )
\end{array}
$
\\[7mm]
\tcolrow $w=-8$
\\[2mm]
\hline
\\[-2mm]
$\frac{\hat E_2^2}{\varDelta} = \frac{1}{2854051200} {\mathcal F} (7,1,-8) - \frac{1}{3991680} {\mathcal F} (5,1,-8)$
\\[4mm]
\tcolrow $w=-10$
\\[2mm]
\hline
\\[-2mm]
$\frac{\hat E_2}{\varDelta} = \frac{1}{13!} {\mathcal F} (7,1,-10)$
\end{tabular}
    \caption{List of all weak almost  holomorphic modular forms of negative weight with
a simple pole at  $q=0$, as  linear combination of Niebur-Poincar\'e series ${\mathcal F} (1-\frac{w}{2} + n, 1, w)$ (the holomorphic ones appear in the first column of Table \ref{holtable}). 
    }
      \label{maintable}
  \end{table}

\subsection{One-loop BPS amplitudes as BPS-state sums}
\label{sec_BPS}

Since any weak almost  holomorphic modular form of negative weight can be represented 
as a linear combination of Niebur-Poincar\'e series,  for the purpose of computing
integrals of the form \eqref{modintegral} it suffices to consider the basic integral
\begin{equation}
{\mathcal I}_{d+k,d} (G,B,Y; s,\kappa ;  \cT) \equiv {\mathcal I}_{d+k,d}  (s,\kappa; \cT )
=\int_{{\mathcal F}_\cT} \de\mu\, \varGamma_{d+k,d} (G,B,Y)\, {\mathcal F} (s, \kappa , -\tfrac{k}{2})\,,
\label{mainint}
\end{equation}
where the modular weight  $w=-k/2$ of the Niebur-Poincar\'e series is determined, via modular invariance, by the signature of the Narain lattice. In order to regulate potential infrared divergences, associated to massless string states, we have introduced in \eqref{mainint} an infrared cut-off $\cT$, which we shall eventually take to infinity.

According to the  unfolding procedure, extended in the presence of a hard cut-off $\cT$ in
\cite{MR656029},  the truncated fundamental 
domain $\cF_{\cT}$ can be extended to the truncated strip $\{ 0<\tau_2<\cT, 
-\tfrac12 \le \tau_1 < \tfrac12\}$ at the expense of restricting the sum over images
in the Niebur-Poincar\'e series  to the trivial coset, and subtracting the 
contribution of the non-trivial ones integrated over the complement $\cF-\cF_\cT$. In equations
\begin{equation}
\begin{split}
{\mathcal I}_{d+k,d} (s,\kappa,\cT ) =& \int_0^{\mathcal T} \frac{\de\tau_2}{\tau_2^2} \,
\int_{-1/2}^{1/2}\, \de\tau_1\, 
 \varGamma_{d+k,d}  \, {\mathcal M}_{s,-\frac{k}{2}} (-\kappa \tau_2 )\, e^{-2\I\pi\kappa \tau_1 }
\\
& - \int_{{\mathcal F} - {\mathcal F}_{\mathcal T}} \de\mu\, \varGamma_{d+k,d}\,
 \left( {\mathcal F} (s, \kappa , -\tfrac{k}{2}) - {\mathcal M}_{s,-\frac{k}{2}} (-\kappa \tau_2 ) \, 
 e^{-2\I\pi\kappa \tau_1 }\right)\, .
\end{split}
\label{I1}
\end{equation}
Using the asymptotic behaviours
\be
 \cM_{s,-\frac{k}{2}}(-\kappa \tau_2) \sim \tau_2^{s+\frac{k}{4}}\,,\quad
  \varGamma_{d+k,d} \sim \tau_2^{-\frac{d+k}{2}}\ ,\qquad {\rm as}\quad \tau_2 \to 0\,,
 \ee 
 and
 \begin{equation}
    \varGamma_{d+k,d} \sim \tau_2^{\frac{d}{2}}\qquad {\rm as}\quad \tau_2 \to\infty\,,
\end{equation}
 together with the Fourier expansion \eqref{FskwF}, one can show that the second integral in \eqref{I1}  
 converges for $\Re(s)>\tfrac14(2d+k)$, while  the first integral in \eqref{I1} converges for 
 $\Re(s)>1+\tfrac14(2d+k)$. For $\Re(s)$ in this range, one may then remove the IR cut-off and extend, 
 in the first integral, the $\tau_2$ range to the full $\IR_+$.
 Moreover, the $\tau_1$ integral vanishes unless the lattice vector satisfies the level-matching constraint
 \be
 \label{levelm}
 p_{\rm L}^2 - p_{\rm R}^2 = 4\kappa\,.
 \ee
In heterotic string vacua (with $\kappa =1$)  this condition selects the contributions of the half-BPS states in the perturbative spectrum, and thus the first integral  in \eqref{I1} can be written as a BPS-state sum 
 \be
 \label{defS2F1}
 {\mathcal I}_{d+k,d}  (s,\kappa) \equiv  \sum_{\rm BPS} \, 
 \int_0^{\mathcal \infty} \frac{\de\tau_2}{\tau_2^2} \, 
 {\mathcal M}_{s,-\frac{k}{2}} (-\kappa \tau_2 )\, \tau_2^{d/2} \, 
e^{-\pi\tau_2 (p^2_{\rm L} + p^2_{\rm R})/2}\,.
 \ee
Here we have introduced the short-hand notation
\begin{equation}
\sum_{\rm BPS} \equiv \sum_{p_{\rm L}\,,\,p_{\rm R}} \, \delta (p_{\rm L}^2 - p_{\rm R}^2 - 4\kappa )
\end{equation}
to denote the sum over those lattice vectors satisfying the level-matching
 condition \eqref{levelm}, and corresponding to half-BPS states if $\kappa =1$. By the previous estimates, this sum is absolutely convergent for 
 $\Re(s)>1+\tfrac14(2d+k)$, and thus defines an analytic function of $s$ in this range.

 To relate  the BPS-state sum to the modular integral of interest,  we note that upon using \eqref{I1} and rearranging terms, Eq. \eqref{defS2F1} may be rewritten as 
  \be
  \label{I2}
 \begin{split}
 {\mathcal I}_{d+k,d}  (s,\kappa) 
= &{\mathcal I}_{d+k,d} (s,\kappa,\cT ) \\ 
&+\int_{{\mathcal F} - {\mathcal F}_{\mathcal T}} \de\mu\, \varGamma_{d+k,d}\,
 \left( {\mathcal F} (s, \kappa , -\tfrac{k}{2}) - {\mathcal M}_{s,-\frac{k}{2}} (-\kappa \tau_2 )\,
 e^{-2\I\pi\kappa \tau_1 }\,
 - f_0(s)\, \tau_2^{1-s+\frac{k}{4}} \right) 
\\
 &+\int_{{\mathcal F} - {\mathcal F}_{\mathcal T}} \de\mu\, (\varGamma_{d+k,d}- \tau_2^{\frac{d}{2}})\,
 \left(  {\mathcal M}_{s,-\frac{k}{2}} (-\kappa \tau_2 ) \, 
 e^{-2\I\pi\kappa \tau_1 } + f_0(s)\, \tau_2^{1-s+\frac{k}{4}} \right)
 \\
 &+\int_{{\mathcal F} - {\mathcal F}_{\mathcal T}} \de\mu\,  \tau_2^{\frac{d}{2}}\,
 \left(  {\mathcal M}_{s,-\frac{k}{2}} (-\kappa \tau_2 ) \, 
 e^{-2\I\pi\kappa \tau_1 } + f_0(s)\, \tau_2^{1-s+\frac{k}{4}} \right)\,,
  \end{split}
 \ee
where 
\begin{equation}
f_0(s)  =\frac{(4\pi)^{1+\frac{k}{4}} \pi^s\, {\I}^{\frac{k}{2}}
\,  \varGamma(2s-1)\, \kappa^{s+\frac{k}{4}} \, \sigma_{1-2s}(\kappa)}
{\varGamma(s+\frac{k}{4}) \, \varGamma(s-\frac{k}{4})\, \zeta(2s)}
\end{equation}
is the coefficient of the zero-frequency Fourier mode \eqref{Fzero}, and the {\em r.h.s.} of \eqref{I2} is independent of $\cT$. The first three lines in 
\eqref{I2} are analytic functions of $s$ for $\Re(s)>1$,  since  ${\mathcal I}_{d+k,d} (s,\kappa , {\mathcal T})$ is integrated over  the compact domain $\cF_\cT$, while the integrands in the second and third
line are exponentially suppressed as $\tau_2\to \infty$, away from the points of enhanced 
gauge symmetry. The fourth line, however, evaluates to  
\be
f_0(s)\, \int_{\cT}^{\infty} \de\tau_2\, \tau_2^{-1-s+\frac{2d+k}{4}}\, 
= f_0(s) \, \frac{{\mathcal T}^{\frac{2d+k}{4}-s}}{s-\frac{2d+k}{4}}
\ee
and is therefore analytic in $s$, except for a simple pole at $s=\tfrac14(2d+k)$. 
We thus conclude that the BPS state sum  \eqref{defS2F1} admits a meromorphic
continuation to $\Re(s)>1$, with a simple pole at $s=\frac{2 d+k}{4}$ with 
residue $f_0(\frac{2d+k}{4})$. Moreover, taking the limit $\cT\to\infty$ in \eqref{I2}, we find
that   the BPS-state sum \eqref{defS2F1}  is actually equal to the renormalised integral 
\be
\label{Irenorm1}
\begin{split}
{\rm R.N.}\,\int_{\cF}\,\de\mu\, \varGamma_{d+k,d}\, {\mathcal F} (s, \kappa , -\tfrac{k}{2}) &=
\lim_{\cT\to\infty} 
 \left[ {\mathcal I}_{d+k,d} (s,\kappa,\cT )  + f_0(s) \frac{{\mathcal T}^{\frac{2d+k}{4}-s}}{s-\frac{2d+k}{4}} \right] 
 \\
 &= {\mathcal I}_{d+k,d}  (s,\kappa) 
\end{split}
\ee
for generic values of  $s\neq \frac{2d+k}{4}$. At the point  $s = \frac{2d+k}{4}$,
the renormalised integral is instead equal to the constant term in the Laurent expansion
of  ${\mathcal I}_{d+k,d}  (s,\kappa)$ around $s= \frac{2d+k}{4}$,
\be
\label{Irenorm2}
\begin{split}
{\rm R.N.}\,\int_{\cF}\,\de\mu\, \varGamma_{d+k,d}\, {\mathcal F} (s, \kappa , -\tfrac{k}{2})  &=
\lim_{\cT\to\infty} 
 \left[ {\mathcal I}_{d+k,d} \left( \tfrac{2 d+k}{4},\kappa,\cT \right)  - f_0\left(\tfrac{2d+k}{4}\right) \, \log\cT +
 f'_0\left(\tfrac{2 d+k}{4}\right) \right]
 \\
&=\hat{\mathcal I}_{d+k,d}  \left( \tfrac{2 d+k}{4},\kappa \right) \,,
\end{split}
\ee
where $f_0'(s)=\de f_0/\de s$, and the {\em r.h.s.} is defined as the 
limit of ${\mathcal I}_{d+k,d} (s,\kappa)$ after the pole is properly subtracted,
\be
\hat{\mathcal I}_{d+k,d}  \left( \frac{2d+k}{4},\kappa\right) 
\equiv \lim_{s\to  \frac{2d+k}{4}}  \left[ {\mathcal I}_{d+k,d} (s,\kappa)  
- \frac{f_0\left(\frac{2d+k}{4}\right)}{s-\frac{2d+k}{4}}
\right] \,.
\ee
Eqs. \eqref{Irenorm1} and \eqref{Irenorm2} relate the renormalised integral to
 the BPS state sum \eqref{defS2F1}, or to its analytic continuation 
whenever $\Re(s)>1$.  We note that this renormalisation prescription 
amounts to subtracting only the infrared divergent contribution of the massless states, 
unlike other schemes used in the literature where the full contribution of the massless states 
is subtracted. Of course, any two renormalisation schemes differ by an additive constant independent of the moduli.

Having discussed the analytic properties of the BPS-state sum \eqref{defS2F1}, and its relation to the regulated integral \eqref{mainint}, let us now  evaluate the integral in \eqref{defS2F1}. Using the relation \eqref{M1F1} between the 
Whittaker $M$-function  and the confluent hypergeometric function $_1F_1$, as well 
as the identity
\begin{equation}
\int_0^\infty dt\, t^{a -1}\, e^{- z\, t}\, {}_1 F_1 (b ; c ; t) = z^{-a} \, \varGamma (a ) \, {}_2 F _1 (a , b ; c ; z^{-1} )\,,
\label{int1F1}
\end{equation}
we  arrive at our main result
\begin{equation}
\begin{split}
{\mathcal I}_{d+k,d} (s,\kappa ) 
=& (4\pi \kappa )^{1-\frac{d}{2}}\, \varGamma (s+ \tfrac{2d+k}{4} -1) 
\\
&\times \sum_{\rm BPS}\,\, {}_2 F_1 \left(s-\frac{k}{4} \,,\, s+ \frac{2d+k}{4} -1 \,;\, 2s \,;\, \frac{4 \kappa}{ p_{\rm L}^2} \right)\, \left(\frac{p_{\rm L}^2}{4\kappa} \right)^{1-s- \frac{2d+k}{4}}  \,.
\end{split}
\label{int2F1}
\end{equation}
The sum in \eqref{int2F1} converges absolutely for 
$\Re(s)>\frac{2d+k}{4}$ and can be analytically continued to a meromorphic function
on $\Re(s)>1$ with a simple pole at $s=\frac{2d+k}{4}$ \cite{1004.11021}. Again, for $\kappa =1$ the sum in 
\eqref{int2F1} can be physically  interpreted as a sum of the one-loop contributions
of all physical BPS states satisfying the level-matching 
condition \eqref{levelm}. This expression is manifestly invariant under T-duality, independent of any choice of chamber, and generalises the 
constrained Epstein zeta series considered in  \cite{Obers:1999um,Angelantonj:2011br} to the case of a non-trivial elliptic genus. We would like to stress that  these properties follow directly from our approach, as opposed to the conventional unfolding method, which depends on a choice of chamber to ensure convergence.

Moreover, using the fact that the lattice partition function satisfies the differential
equation \cite{Obers:1999um}
\begin{equation}
\label{DelSOG}
\left[ \Delta_{{\rm SO} (d+k,d)} -2\, \Delta_{k/2} + \tfrac14 \,d(d+k-2) \right]\, \varGamma_{d+k,d} = 0\,,
\end{equation} 
we find that the BPS state sum \eqref{int2F1} is an eigenmode of the Laplacian $\Delta_{{\rm SO} (d+k,d)}$  on the Narain moduli space
\begin{equation}
\label{LapsEpstein}
\begin{split}
\left[ \Delta_{{\rm SO} (d+k,d)}+\tfrac{1}{16}\, (2d+k-4s)(2d+k+4s-4) \right]\, 
{\mathcal I}_{d+k,d} (s,\kappa )  =0\ .
\end{split}
\end{equation}
For $s=\frac{2d+k}{4}$, the eigenvalue vanishes but the BPS state sum ${\mathcal I}_{d+k,d} (s,\kappa)$ has a pole. After subtracting the pole, one finds that the renormalised BPS state sum 
is an almost harmonic function on the Narain moduli space, namely its image under the Laplacian
is a constant
\begin{equation}
\label{LapsEpstein2}
\begin{split}
 \Delta_{{\rm SO} (d+k,d)} \, 
\hat {\mathcal I}_{d+k,d} \left(\frac{2d+k}{4},\kappa \right)  =(1-d-\tfrac{k}{2})\, f_0 \left(\frac{2d+k}{4}\right)\,.
\end{split}
\ee

\subsection{One-loop BPS amplitudes with momentum insertions}

Our method carries over straightforwardly to cases where insertions of left-moving  
or right-moving momenta appear  in the lattice sum, i.e. to modular integrals of the type 
\begin{equation}
\label{intinsert}
\int_{\mathcal F} \de \mu \, \left[ \tau_2^{-\lambda/2} \, \sum_{p_{\rm L}, p_{\rm R}} 
\rho\left(p_{\rm L}\sqrt{\tau_2},p_{\rm R}\sqrt{\tau_2}\right) \, 
q^{\frac{1}{4} p_{\rm L}^2}\, \bar q^{\frac{1}{4} p_{\rm R}^2}\right] \, \varPhi (\tau)\,,
\end{equation}
considered for example in  \cite{Lerche:1999ju, 0919.11036}. The term in the square
bracket is a modular form of weight $(\lambda+d+\tfrac{k}{2},0)$, provided that
the function $\rho(x_{\rm L}\,, \,x_{\rm R})$ satisfies 
\be
\label{EqVigneras}
\left[ \partial_{x_{\rm L}}^2-\partial_{x_{\rm R}}^2 - 2\pi \left(x_{\rm L} \partial_{x_{\rm L}} - x_{\rm R} \, \partial_{x_{\rm R}} -\lambda-d\right) \right]\, \rho(x_{\rm L} \,,\, x_{\rm R}) = 0\ ,
\ee
and that $\rho(x_{\rm L},\, x_{\rm R} ) \, e^{-\frac{\pi}{2} (x_{\rm L}^2+x_{\rm R}^2)}$ should decay sufficiently fast at infinity \cite{Vigneras}\footnote{We are grateful to J. Manschot for pointing out 
this reference.}. For example, upon choosing $(\rho=x_{\rm L}^2-\frac{d+k}{2\pi},\lambda=2-d)$
(respectively $(\rho=x_{\rm R}^2-\frac{d}{2\pi},\lambda=-2-d)$),
it is proportional to the modular derivative 
$D\cdot \varGamma_{d+k,d}$ (respectively,   $\overline{D}\cdot \varGamma_{d+k,d}$)
of the usual Narain lattice partition function. The integrand in  \eqref{intinsert} is then modular invariant provided $\lambda+d+\tfrac{k}{2}=-w$.

As usual, expressing the elliptic genus as a linear combination of Niebur Poincar\'e series, one is left to consider integrals of the form
\begin{equation}
\int_{\mathcal F} \de \mu \, \tau_2^{-\lambda/2} \, \sum_{p_{\rm L}, p_{\rm R}} 
\rho\left(p_L\sqrt{\tau_2},p_R\sqrt{\tau_2}\right) \, 
q^{\frac{1}{4} p_{\rm L}^2}\, \bar q^{\frac{1}{4} p_{\rm R}^2}\, {\mathcal F} (s, \kappa , w)\,,
\end{equation}
and following similar steps as in the previous subsection, one finds the result
\begin{equation}
(4\pi\kappa )^{1+\frac{\lambda}{2}}\, \sum_{\rm BPS} \int_0^\infty dt\, t^{s + \frac{2d+k}{4}-2}\, {}_1 F_1 \left( s - \frac{2\lambda + 2 d + k}{4}; 2s;t \right)\, \rho \left( \frac{p_{\rm L}}{\sqrt{4\pi\kappa}}\, \sqrt{t} , 
 \frac{p_{\rm R}}{\sqrt{4\pi\kappa}}\, \sqrt{t} \right)\, e^{-t\, p_{\rm L}^2 / 4\kappa}\,.
 \end{equation}
In  most applications, $\rho$ is a polynomial in $p^a_{\rm L}, p^b_{\rm R}$, 
and the integral  can be evaluated using  \eqref{int1F1}. As a result, each monomial can be evaluated to
\be
\begin{split}
\int_{\mathcal F} \de \mu \, \tau_2^\delta\, \sum_{p_{\rm L}, p_{\rm R}} & p_{\rm L}^{a_1} \cdots p_{\rm L}^{a_\alpha} 
 \, p_{\rm R}^{b_1} \cdots p_{\rm R}^{b_\beta} \, q^{\frac{1}{4} p_{\rm L}^2} \bar q^{\frac{1}{4} p_{\rm R}^2} \, {\mathcal F} (s,\kappa , w) \rightrightarrows
  (4\pi \kappa )^{1-\delta}\, \varGamma (s+\tfrac{|w|}{2}+\delta-1)
\\
\times&\sum_{\rm BPS} p_{\rm L}^{a_1} \cdots p_{\rm L}^{a_\alpha}
 \, p_{\rm R}^{b_1} \cdots p_{\rm R}^{b_\beta}\, {}_2 F_1 \left(s- \frac{|w|}{2} , s+\frac{|w|}{2} +\delta -1;2s;\frac{4\kappa}{p_{\rm L}^2} \right) \left(\frac{p_L^2}{4\kappa}\right)^{1-s-\frac{|w|}{2}-\delta}\, ,
\end{split}
\end{equation}
with $\delta=(\alpha+\beta-\lambda)/2$.
Clearly, this result is meaningful only when the various monomials are combined into a solution of \eqref{EqVigneras}, as required by modular invariance.

\subsection{BPS-state sum for integer $s$}

For special values of $s$ and $w$, the hypergeometric function ${}_2 F_1$ appearing in the BPS-state sum \eqref{int2F1} can actually be expressed in terms of elementary functions. For example, for $d=1$ and $w=0$, ${}_2 F_1(s,s-\tfrac12,2s;z)= 2^{2s-1} (1+\sqrt{1-z})^{1-2s}$, and thus
\begin{equation}
\begin{split}
{\mathcal I}_{1,1} (1+n,\kappa) &= \sqrt{4 \pi \kappa}\, 2^{1+2n}\, \varGamma ( n + \tfrac{1}{2} ) \, \sum_{\rm BPS}\,\,\left( \sqrt{\frac{p^2_{\rm L}}{4\kappa} } + \sqrt{\frac{p^2_{\rm R}}{4\kappa}} \right)^{-1-2n}
\\
&= \tfrac{1}{2}\, \sqrt{\pi}\,(16\, \kappa)^{1+ n}\,\varGamma (n+\tfrac{1}{2} ) \, 
\sum\limits_{\substack{ p,q\in\mathbb{Z}\\ pq=\kappa }}
\left( \left| p\, R + q\, R^{-1} \right| + \left| p\, R - q\, R^{-1} \right| \right)^{-1-2n}\,,
\label{explicitd1}
\end{split}
\end{equation}
with $s=1+n$.
For $n=0$, this agrees with the expression derived in \cite{Angelantonj:2011br}
using the Selberg-Poincar\'e series $\cE(s,\kappa,w)$ at $s=0$.

More generally, similar simplifications also take place for $s=1-\tfrac{w}{2}+n=1+\frac{k}{4}+n$, with $n$ a positive integer, which are the special values relevant for representing weak almost  holomorphic modular forms, and are thus of interest for our physical applications. While it is cumbersome to express
 ${}_2 F_1$ directly in terms of elementary functions, it is  simpler to notice that the Whittaker $M$-function appearing in \eqref{defS2F1} reduces to the finite sum \eqref{Mn}. As a result, the integral \eqref{defS2F1} reduces to 
\begin{equation}
\label{Irenorma}
\begin{split}
{\mathcal I}_{d+k,d} (1+\tfrac{k}{4}+n,\kappa)   =& \sum_{\rm BPS}\,\,
\int_0^\infty \de \tau_2 \, \tau_2^{\frac{d}{2}-2+\alpha} \, {\mathcal M}_{1+\frac{k}{4}+n,-\frac{k}{2}} (-\kappa \tau_2 ) \,  e^{-\pi \tau_2 (p_{\rm L}^2 + p_{\rm R}^2 )/2} \,,
\\
=& 
 \, (4 \pi \kappa )^{1-\tfrac{d}{2}}\,  \frac{\varGamma (2 (n+1) +\tfrac{k}{2} ) \, 
 \varGamma (n + \tfrac{d+k}{2})}{n!} \sum_{m=0}^n \binom{n}{m} \frac{(-1)^m }
 {\varGamma (n-m +\tfrac{d+k}{2})}
\\
&\times \sum_{\rm BPS}\, \left( \frac{p_{\rm L}^2 }{4\kappa}\right)^{n-m}
\int_0^\infty \de z \,\, z^{\tfrac{d}{2}-m-2} \left( e^{-z \, p_{\rm R}^2 / 4\kappa} -
e^{-z \, p_{\rm L}^2 / 4\kappa}  
\sum_{\ell =0}^{2n + \tfrac{k}{2}} \frac{z^\ell}{\ell !} \right) \, ,
\end{split} 
\end{equation}
where, in going from the first to the second line we have set $z=4\pi \kappa \tau_2$, and  have integrated by parts $n$ times,  and we used the fact that the boundary terms vanish.
Although the full integrand  vanishes rapidly enough as $z\to 0$, so that the integral exists, this is not true of each individual term, unless $\frac{d}{2}-n-1 > 0$. To regulate  these unphysical divergences, we introduce a convergence factor $\tau_2^\alpha$ in the integrand, and evaluate each integral in \eqref{Irenorma} for large enough $\alpha$.
The desired result is then expressed as the limit
\begin{equation}
\begin{split}
{\mathcal I}_{d+k,d} (1+\tfrac{k}{4}+n,\kappa)  =& (4\pi \kappa )^{1-\frac{d}{2}}\,\frac{\varGamma (2 (n+1) +\tfrac{k}{2} ) \, \varGamma (n + \tfrac{d+k}{2})}{n!} 
\sum_{m=0}^n \binom{n}{m} \frac{(-1)^m }{\varGamma (n-m +\tfrac{d+k}{2})}\, 
\\
&\times \sum_{\rm BPS}\, \left( \frac{p_{\rm L}^2 }{4\kappa}\right)^{n-m}
\lim_{\alpha \to 0}  \left[ \varGamma \left( \tfrac{d}{2} - m -1 +\alpha \right) \, \left( \frac{p_{\rm R}^2}{4\kappa } \right)^{m+1-\frac{d}{2}-\alpha} \right.
\\
& \qquad\qquad \qquad\left. - \sum_{\ell=0}^{2n + k/2} \frac{\varGamma \left( \frac{d}{2} - m -1 +\ell +\alpha \right)}{\ell !} \left( \frac{p_{\rm L}^2}{4\kappa} \right)^{1+m-\frac{d}{2}-\ell -\alpha} \right] \, .
\end{split}
\label{limit}
\end{equation}
The series \eqref{limit} converges absolutely for $n>\tfrac{d}{2}-1$, as a result of the finiteness of the original modular integral. For $n\leq \frac{d}{2}-1$, it is a formal (divergent) sum over BPS states, which nevertheless  captures the singularities of the amplitude at points of gauge symmetry enhancement.

For $n < \frac{d}{2} -1$, or whenever $d$ is odd, independently of $n$, the limit $\alpha \to 0$ is trivial, leading to ${\mathcal I}_{d+k,d} (s,\kappa ) 
= {\mathcal I}^{(1)}_{d+k,d} (s,\kappa ) $ where
\begin{equation}
\begin{split}
{\mathcal I}^{(1)}_{d+k,d} (1+\tfrac{k}{4}+n,\kappa ) &  = (4\pi \kappa )^{1-\frac{d}{2}}\,\frac{\varGamma (2 (n+1) +\frac{k}{2} ) \, \varGamma (n + \frac{d+k}{2})}{n!}
\\
&\times \, \sum_{m=0}^{d/2-2} \binom{n}{m} \frac{(-1)^m }{\varGamma (n-m +\frac{d+k}{2} )}\, \sum_{\rm BPS}\, \left( \frac{p_{\rm L}^2 }{4\kappa}\right)^{n-m}
\\
&\times  \Biggl[ \varGamma \left( \tfrac{d}{2} - m -1  \right) \, \left( \frac{p_{\rm R}^2}{4\kappa } \right)^{m+1-\frac{d}{2} } 
\\
&\qquad \qquad \qquad
- \sum_{\ell=0}^{2n + k/2} \frac{\varGamma \left( \frac{d}{2} - m -1 +\ell  \right)}{\ell !} \left( \frac{p_{\rm L}^2}{4\kappa} \right)^{1+m-\frac{d}{2}-\ell } \Biggr] \,.
\end{split}
\label{integral1}
\end{equation}
If $d$ is even and $n \ge \frac{d}{2} -1$ one finds
${\mathcal I}_{d+k,d} (s,\kappa ) = {\mathcal I}^{(1)}_{d+k,d} (s,\kappa ) 
+ {\mathcal I}^{(2)}_{d+k,d} (s,\kappa ) $ where the first term is still given by \eqref{integral1}
and the second term is 
\begin{equation}
\begin{split}
{\mathcal I}^{(2)}_{d+k,d} (1+\tfrac{k}{4}+n,\kappa )  &=  (4\pi \kappa )^{1-\frac{d}{2}}\, \frac{\varGamma (2 (n+1) +\frac{k}{2} ) \, \varGamma (n + \tfrac{d+k}{2})}{n!}
\\
&\times  \sum_{\rm BPS}\,\, \sum_{m=d/2-1}^{n} \binom{n}{m} \frac{(-1)^m }{\varGamma (n-m +\tfrac{d+k}{2})}\, \left( \frac{p_{\rm L}^2 }{4\kappa}\right)^{n-m}
\\
&\times \left\{  - \sum_{\ell=m+2-d/2}^{2n + k/2} \frac{\varGamma \left( \tfrac{d}{2} - m -1 +\ell  \right)}{\ell !} \left( \frac{p_{\rm L}^2}{4\kappa} \right)^{1+m-\tfrac{d}{2}-\ell } \right.
\\
&\quad + \frac{(-1)^{m+1-\frac{d}{2}}}{\varGamma (m+2 -\frac{d}{2} )} \left( \frac{p_{\rm R}^2}{4\kappa} \right)^{m+1-\frac{d}{2}} \, \left[ H_{m+1-\frac{d}{2}} - \log \, \left( \frac{p_{\rm R}^2}{p_{\rm L}^2} \right) \right]
\\
&\quad \left. - \frac{1}{\varGamma (m+2-\frac{d}{2} )}\sum_{\ell=0}^{m+1-d/2} \binom{m+1-\frac{d}{2}}{\ell} \, \left( -\frac{p_{\rm L}^2}{4\kappa } \right)^{m+1-\frac{d}{2} -\ell}\, H_{m+1-\frac{d}{2}-\ell} \right\},
\end{split}
\label{integral2}
\end{equation} 
where $H_N = \sum_{k=1}^N k^{-1}$ is the $N$-th harmonic number.
The combination \eqref{integral1} vanishes for $d=2$,  the sum over $m$ being void. 
The results \eqref{integral1} and \eqref{integral2} allow to write any integral of the type \eqref{modintegral} as a formal sum over physical 
BPS states (which converges absolutely for $n>\tfrac{d}{2}-1$). 
In particular, the result is manifestly invariant 
under the T-duality group ${\rm O} (d+k,d;\mathbb{Z})$

We conclude this subsection with some simple examples for special values of $n$ and $k$. For  
$n=0$ the sum over $m$ in \eqref{Irenorma} is void and only few terms contribute to the integral, corresponding to the various terms in \eqref{Mharm}. 
When $d\neq 2$ the limit $\alpha \to 0$ is trivial and one arrives at the simple expression
\be
\begin{split}
{\mathcal I}_{d+k,d} (1+\tfrac{k}{4},\kappa ) = (4\pi\kappa )^{1-\frac{d}{2}}\, \varGamma (2+\tfrac{k}{2} )\, &\sum_{\rm BPS} \Biggl[ \varGamma (\tfrac{d}{2}-1 )\, \left( \frac{p_{\rm R}^2 }{4\kappa }\right)^{1-\frac{d}{2}}
\\
&\qquad\qquad - \sum_{\ell =0}^{k/2} \frac{\varGamma (\frac{d}{2} +\ell -1)}{\ell !} \left( \frac{p_{\rm L}^2}{4\kappa}\right)^{1-\frac{d}{2}-\ell}\Biggr]\,.
\end{split}
\ee 
When $d=2$, the limit $\alpha \to 0$ is subtler and leads to logarithmic contributions. One obtains, for $n=0$,  any $k$,
\be
{\mathcal I}_{2+k,2} (1+\tfrac{k}{4},\kappa)= - \varGamma (2+\tfrac{k}{2})\,
\sum_{\rm BPS} \, \left[\log \left( \frac{p_{\rm R}^2}{p_{\rm L}^2} \right)+\sum_{\ell=1}^{k/2} \,\frac{1}{\ell}\, \left(\frac{p_{\rm L}^2}{4\kappa}\right)^{-\ell}\right] \,,
\ee
and for $k=0$,  any $n$,
\begin{equation}
\begin{split}
{\mathcal I}_{2,2} (1+n,\kappa) &=  \frac{(2n+1)!}{n!}\, \sum_{\rm BPS}\, \, \left( \frac{p_{\rm L}^2}{4\kappa} \right)^n \, \sum_{m=0}^n \binom{n}{m}^2 \left\{ \left(\frac{p_{\rm R}^2}{p_{\rm L}^2} \right)^m \left[ H_m - \log \left(\frac{p_{\rm R}^2}{p_{\rm L}^2} \right) \right]\right.
\\
&\left. - \sum_{\ell =0}^m (-1)^\ell\, \binom{m}{\ell} \, \left( \frac{p_{\rm L}^2}{4\kappa}\right)^{-\ell}\, H_{m-\ell}
- (-1)^m\,\sum_{\ell =m+1}^{2n} \frac{\varGamma (\ell-m) \, m!}{\ell !} \, \left( \frac{p_{\rm L}^2}{4\kappa} \right)^{-\ell} \right\} \,.
\end{split}
\label{d2general}
\end{equation}
As usual, the left and right-handed momenta are defined by 
\begin{equation}
\begin{split}
p_{{\rm L},I} &= ( m_i + Y^a_i Q^a +\tfrac{1}{2} Y^a_i Y^a_j n^j + (G + B)_{ij} n^j ~,~ Q^a + Y^a_j n^j ) \, ,
\\
p_{{\rm R},i} &=  m_i + Y^a_i Q^a +\tfrac{1}{2} Y^a_i Y^a_j n^j - (G - B)_{ij} n^j \, ,
\end{split}
\end{equation}
with $m_i$ and $n^i$ the Kaluza-Klein and winding numbers and $Q^a$ are the charge vectors. In the $d=2$ case, and in the absence of Wilson lines, it is often convenient to express them in terms  of the K\"ahler  modulus $T$  and of the complex structure modulus $U$ as
\begin{equation}
\begin{split}
p_{\rm L}^2 &= \frac{1}{T_2\, U_2} \left| m_2 - U\, m_1 +\bar  T (n^1 + U\, n^2 ) \right|^2 \,,
\\
p_{\rm R}^2 &= \frac{1}{T_2\, U_2} \left| m_2 - U\, m_1 + T (n^1 + U\, n^2 ) \right|^2 \,.
\end{split}
\end{equation}

The relation between these results (for $k=0$) and the `shifted constrained Epstein zeta series'
of \cite{Angelantonj:2011br} is discussed in Appendix \ref{sec_Selberg}.

\subsection{Singularities at points of  gauge symmetry enhancement\label{sec_sing}}

In addition to keeping T-duality manifest, another advantage of this approach for the evaluation of one-loop modular integrals is that it allows to easily read-off the singularity structure of the amplitudes 
at point of  enhanced gauge symmetry. These points are characterised by the appearance of extra massless states with $p_{\rm R} =0$. Depending on the dimension of the Narain lattice, as well as on the value of $n$, the amplitude may diverge ( we refer to this case  as real singularity) or one of its derivatives can be discontinuous (we refer to this case as conical singularity).

For odd dimension $d$, the modular integral ${\mathcal I}_{d+k,d} (s,\kappa)$ always develops {\em conical} singularities, as exemplified in the one-dimensional case by Eq. \eqref{explicitd1}. In addition, 
for $d\ge 3$ real singularities  appear from terms with $m<\frac{d}{2} -1$ in \eqref{limit}.

For even dimension real singularities always appear. They are are power-like in ${\mathcal I}^{(1)}$ whenever $d\ge 4$ and logarithmic in ${\mathcal I}^{(2)}$ for any even $d \le 2n+2$. Moreover, conical singularities do not appear. 

Notice that for $d=2$ the singularities cancel in the combination
\begin{equation}
{\mathcal I}_{2,2} (1+n,1) - \frac{(2n+1)!}{n!} \,\, \hat{\mathcal I}_{2,2} (1,1)  \, ,
\label{singular}
\end{equation}
which is therefore a continuous function over the Narain moduli space, including at points of
enhanced gauge symmetry. Since, using the results in \cite{Dixon:1990pc,Harvey:1995fq} and the fact that ${\mathcal F} (1,1,0) = j + 24$,
\begin{equation}
\hat{\mathcal I}_{2,2} (1,1) = - \log | j(T) - j(U) |^4  - 24 \, \log \left[ T_2 U_2 |\eta (T) \, \eta (U) |^4 \right] + {\rm const}\,,
\end{equation}
we  conclude that all integrals ${\mathcal I}_{2,2} (1+n,1)$ exhibit the same universal singular behaviour up to an overall normalisation
\begin{equation}
{\mathcal I}_{2,2} (1+n,1) \sim -\frac{(2n+1)!}{n!}\, \log | j(T) - j(U) |^4 \,.
\label{singularity}
\end{equation}
This expression can be generalised as in \cite{Harvey:1995fq} if Wilson lines are turned on.

 \section{Some examples from string threshold computations}\label{SecExamples}
 
 In this section we evaluate a sample of modular integrals that enter in  threshold corrections to gauge and gravitational couplings in heterotic string vacua using the method developed in the previous section.
We express the elliptic genus as a linear combination of Niebur-Poincar\'e series, and we evaluate the modular integral in terms of the BPS-state sums ${\mathcal I}_{d+k,d} (s,\kappa )$ defined in eqs. \eqref{int2F1} and \eqref{limit}.

 \subsection{A gravitational coupling in maximally supersymmetric heterotic vacua}
 
 Let us start with the example of toroidally compactified ${\rm SO} (32) $ heterotic string, for which the elliptic genus takes the form \eqref{N4thresholds}. Using Table \ref{maintable} and the relation $E_4^3 \, \varDelta^{-1} = j + 744$, this can be conveniently expressed in terms of the Niebur-Poincar\'e series as
 \begin{equation}
 \begin{split}
 \varPhi (\tau ) =& t_8 \, {\rm tr}\, F^4 + \frac{1}{2^7\, 3^2 \, 5} \left[ {\mathcal F} (1,1,0) + 720\right]\, t_8 \, {\rm tr} \, R^4 
 \\
 &+ \frac{1}{2^9\, 3^2} \left[ \tfrac{1}{5} {\mathcal F} (3,1,0) - 4 {\mathcal F} (2,1,0) + 13 {\mathcal F} (1,1,0) + 144 \right]\,  t_8\, ({\rm tr}\, R^2 )^2
 \\
 &+ \frac{1}{2^8\, 3^2} \left[ -\tfrac{1}{5} {\mathcal F} (3,1,0) +5 {\mathcal F} (2,1,0) -18 {\mathcal F} (1,1,0) +288 \right] \, t_8\,{\rm tr}\, R^2 \, {\rm tr}\, F^2
 \\ 
 & + \frac{1}{2^9\, 3^2} \left[ \tfrac{1}{5} {\mathcal F} (3,1,0) -6 {\mathcal F} (2,1,0) + 24 {\mathcal F} (1,1,0) -576\right]\, t_8 \, ({\rm tr} \, F^2)^2\,.
\end{split}
\end{equation}
Therefore, using the results in the previous section,  the renormalised modular integral \eqref{modintegral} 
can be expressed as the linear combination  
\begin{equation}
\begin{split}
{\rm R.N.}\int_{\mathcal F} \de\mu \, \varGamma_{d,d} \, \varPhi =&\,
{\mathcal I}_{d,d} \,  t_8 \, {\rm tr}\, F^4 + \frac{1}{2^7\, 3^2 \, 5} \left[ {\mathcal I}_{d,d} (1,1) + 720\, {\mathcal I}_{d,d} \right]\, t_8 \, {\rm tr} \, R^4 
 \\
 &+ \frac{1}{2^9\, 3^2} \left[ \tfrac{1}{5}\, {\mathcal I}_{d,d} (3,1) - 4 \,{\mathcal I}_{d,d} (2,1) + 13 \,{\mathcal I}_{d,d} (1,1) + 144 \, {\mathcal I}_{d,d} \right]\,  t_8\, ({\rm tr}\, R^2 )^2
 \\
 &+ \frac{1}{2^8\, 3^2} \left[ -\tfrac{1}{5}\, {\mathcal I}_{d,d} (3,1) +5\, {\mathcal I}_{d,d} (2,1) -18\, {\mathcal I}_{d,d} (1,1) +288\, {\mathcal I}_{d,d} \right] \, t_8\,{\rm tr}\, R^2 \, {\rm tr}\, F^2
 \\ 
 & + \frac{1}{2^9\, 3^2} \left[ \tfrac{1}{5} \, {\mathcal I}_{d,d} (3,1) -6\, {\mathcal I}_{d,d} (2,1) + 24 \,{\mathcal I}_{d,d} (1,1) -576\, {\mathcal I}_{d,d}\right] \, t_8 \, ({\rm tr} \, F^2)^2\,,
\end{split}
\label{threshgeneral}
\end{equation}
where, as computed in \cite{Obers:1999um,Angelantonj:2011br},
\begin{equation}
{\mathcal I}_{d,d} \equiv {\rm R.N.}\,  \int_{\mathcal F} \de\mu \, \varGamma_{d,d} (G,B) = \frac{\varGamma (\tfrac{d}{2} -1 )}{\pi^{\tfrac{d}{2}-1}}\, {\mathcal E}_V^d (G,B;\tfrac{d}{2} -1 )\,,
\end{equation}
with ${\mathcal E}_V^d$ being the constrained Epstein zeta function defined in \cite{Obers:1999um,Angelantonj:2011br}. In this expression, any time $n=\frac{d}{2}-1$ the BPS-state sum ${\mathcal I}_{d,d} (1+n,\kappa )$ should be replaced by $\hat{\mathcal I}_{d,d} (1+n,\kappa)$, as explained in Section \ref{sec_BPS}. 

In the one-dimensional case the constrained sums can be easily evaluated, leading to
\begin{equation}
\begin{split}
\int_{\mathcal F} \de\mu \, \varGamma_{1,1} \, \varPhi =& 
\frac{\pi}{3} (R+R^{-1}) \,  t_8 \, {\rm tr}\, F^4 
+ \frac{\pi}{2^3\, 3^2 \, 5} \left( 15 R + 16 R^{-1}
 \right)\, t_8 \, {\rm tr} \, R^4 
\\
&+ \frac{\pi}{2^5\, 3^2} \left( 3 R+16 R^{-1} - 24 R^{-3} + 12 R^{-5} \right)\,  t_8\, ({\rm tr}\, R^2 )^2
\\
&+ \frac{\pi}{2^3\, 3} \left( R -2R^{-1} + 5 R^{-3} - 2 R^{-5} \right) \, t_8\,{\rm tr}\, R^2 \, {\rm tr}\, F^2
\\
&-\frac{\pi}{2^3\, 3} \left( R-R^{-1}+3 R^{-3}-R^{-5}\right) \, t_8 \, ({\rm tr} \, F^2)^2\,,
\end{split}
\end{equation}
for $R>1$. The expression for $R<1$ can be obtained by replacing in the previous expression $R\mapsto R^{-1}$. Notice that, aside from the threshold correction to $t_8 \, {\rm tr}\, F^4$, all other terms develop a conical singularity at the self-dual radius $R=1$.

\subsection{Gauge-thresholds in ${\mathcal N}=2$ heterotic vacua with/without Wilson lines}

Let us turn now to ${\mathcal N}=2$ heterotic vacua in the orbifold limit $T^2 \times T^4/{\mathbb Z}_2$, with a standard embedding on the gauge sector. At the orbifold point, the gauge group is  broken to
\begin{equation}
{\rm E}_8 \times {\rm E}_8 \to {\rm E}_8 \times {\rm E}_7 \times {\rm SU} (2)\,,
\end{equation}
and, in the absence of Wilson lines, gauge threshold corrections read
\begin{equation}
\begin{split}
\varDelta_{{\rm E}_8} &= -\frac{1}{12} \int_{\mathcal F}\de \mu \, \varGamma_{2,2}\, \frac{\hat E_2 \, E_4 \, E_6 - E_6^2}{\varDelta}\,,
\\
\varDelta_{{\rm E}_7} &= -\frac{1}{12} \int_{\mathcal F}\de \mu \, \varGamma_{2,2}\, \frac{\hat E_2 \, E_4 \, E_6 - E_4^3}{\varDelta}\,.
\end{split}
\end{equation}
From Table \ref{maintable} one can read that
\begin{equation}
\begin{split}
\frac{\hat E_2 \, E_4 \, E_6 - E_6^2}{\varDelta} &= {\mathcal F} (2,1,0 ) - 6 \, {\mathcal F} (1,1,0) + 864\,,
\\
\frac{\hat E_2 \, E_4 \, E_6 - E_4^3}{\varDelta} &= {\mathcal F} (2,1,0 ) - 6 \, {\mathcal F} (1,1,0) - 864\,,
\end{split}
\end{equation}
and thus
\begin{equation}
\begin{split}
\varDelta_{{\rm E}_8} &=  \sum_{\rm BPS}\,
\left[ 1+\frac{p_{\rm R}^2}{4}\, \log\, \left( \frac{p_{\rm R}^2}{p_{\rm L}^2} \right) \right] + 72\, \log\, \left( T_2\, U_2\, \left|\eta (T) \, \eta (U) \right|^4 \right) +{\rm const}\,,
\\
\varDelta_{{\rm E}_7} &= \sum_{\rm BPS}\,
\left[ 1+\frac{p_{\rm R}^2}{4}\, \log\, \left( \frac{p_{\rm R}^2}{p_{\rm L}^2} \right) \right] - 72\, \log\, \left( T_2\, U_2\, \left|\eta (T) \, \eta (U) \right|^4 \right) +{\rm const}
\end{split}
\end{equation}
Notice that the combination ${\mathcal I}_2 (2,1,0) - 6 \, {\mathcal I}_2  (1,1,0)$ is regular at any point in moduli space  (and in any chamber), as expected since the unphysical tachyon is neutral and therefore does not contribute to the running of the non-Abelian gauge couplings. 

Turning on Wilson lines on the ${\rm E}_8$ group factor along the spectator $T^2$, yields
\begin{equation}
\varDelta_{{\rm E}_7} = -\frac{1}{12}\int_{\mathcal F}\de\mu\, \varGamma_{2,10}\, \frac{\hat E_2\, E_6 - E_4^2}{\varDelta}\,.
\end{equation}
Using Table \ref{maintable}, one easily finds
\begin{equation}
\frac{\hat E_2\, E_6 - E_4^2}{\varDelta} = \frac{2}{7!} {\mathcal F} (4,1,-4) - \frac{2}{5!}\, {\mathcal F} (3,1,-4)\,,
\end{equation}
and thus
\begin{equation}
\begin{split}
\varDelta_{{\rm E}_7} &= -\frac{1}{720}\left[ \frac{1}{42} \, {\mathcal I}_{10,2} (4,1) - {\mathcal I}_{10,2} (3,1) \right] 
\\
&= \sum_{\rm BPS} \, \left[ 1 + \frac{p_{\rm R}^2}{4}\, \log \left( \frac{p_{\rm R}^2}{p_{\rm L}^2}\right) - \frac{2}{p_{\rm L}^2} - \frac{8}{3\, p_{\rm L}^4} -\frac{16}{3\, p_{\rm L}^6}-\frac{64}{5\, p_{\rm L}^8}\right]\,.
\end{split}
\end{equation}
In this expression $p_{\rm L,R}$ depend also on the Wilson lines, and the constraint in the BPS-sum now reads
\begin{equation}
p_{\rm L}^2 - p_{\rm R}^2 =4 \quad \Rightarrow\quad m^{\rm T} n + \tfrac{1}{2} \, Q^{\rm T} Q =1\,,
\end{equation}
where $Q$ is the ${\rm U}(1)$-charge vector in the Cartan sub-algebra of ${\rm E}_8$.

\subsection{K\"ahler metric corrections in ${\mathcal N}=2$ heterotic vacua}

Our procedure can also be used to compute loop corrections to K\"ahler metric and other terms in the low-energy effective action. For instance, in ${\mathcal N}=2$ heterotic vacua at the orbifold point, the one-loop correction to the K\"ahler metric for the $T$ modulus reads
\begin{equation}
\begin{split}
 K_{T\bar T} \Bigr|_{1-{\rm loop}}&= \frac{\I}{12\pi\, T^2_2}  \,\int_{\mathcal F} \de\mu \, \frac{E_4\, E_6}{\varDelta} \, \partial_{\tau} \, \varGamma_{2,2}
 \\
 &= \frac{\I}{72\pi\, T^2_2}  \,\int_{\mathcal F} \de\mu \, {\mathcal F} (2,1,-2) \, \partial_{\tau} \, \varGamma_{2,2}\,,
 \end{split}
\end{equation}
where we have used the relation between $E_4 E_6 \varDelta^{-1}$ and ${\mathcal F} (s,\kappa ,w)$ from Table \ref{holtable}. Integrating by parts, and using the action of the modular derivative on the Niebur-Poincar\'e series, one immediately finds
\begin{equation}
\begin{split}
K_{T\bar T} \Bigr|_{1-{\rm loop}} &= - \frac{\I}{72\pi\, T_2^2} \,\int_{\mathcal F}\de\mu \,\varGamma_{2,2} \, D_{-2} {\mathcal F} (2,1,-2 )
\\
&= \frac{1}{36 \, T_2^2} \, \int_{\mathcal F} \de\mu\, \varGamma_{2,2}\, {\mathcal F} (2,1,0)
\\
&=  \frac{1}{36 \, T_2^2}\, {\mathcal I}_{2,2} (2,1)\,.
\end{split}
\end{equation}
Similar results can be obtained for higher-derivative couplings in ${\mathcal N} =4$ vacua.

\subsection{An example from non-compact heterotic vacua}
 
 In some   heterotic constructions on ALE spaces and in the presence of background NS5 branes, gauge threshold corrections include a contribution of the (finite) integral \cite{luca} 
 \begin{equation}
L= \int_{\mathcal F} \de\mu \, (\sqrt{\tau_2} \, \eta \, \bar\eta )^3 \, \frac{\hat E_2 \, E_4 \left( \hat E_2 \, E_4 - 2 E_6 \right)}{\varDelta} \,.
 \end{equation}
 Despite its  apparent complexity, this integral can be easily computed using our techniques. In fact, the relation
\begin{equation}
\vartheta_1^\prime (0|\tau ) = 2\pi \eta^3\,,
\end{equation}
and the standard bosonisation formulae, allow one to write
\begin{equation}
\left( \sqrt{\tau_2} \, \eta \bar \eta \right)^3 = -\frac{1}{8\pi}\, \frac{\partial}{\partial R} \left[ \frac{1}{R} \left( \varGamma_{1,1} (2R) - \varGamma_{1,1} (R) \right) \right]_{R=1/\sqrt{2}}
\,.
\end{equation}
Combining this observation with Table \ref{maintable}, Eq. \eqref{explicitd1}, and with 
the standard result  $\int_{\mathcal F} \de\mu \, \varGamma_{1,1} (R) = \frac{\pi}{3} (R+R^{-1})$,
the integral reduces to
\begin{equation}
\begin{split}
L &= - \frac{1}{8\pi}\, \frac{\partial}{\partial R} \Biggl[ \frac{1}{R}\, \int_{\mathcal F} \de\mu\, \left( \varGamma_{1,1} (2R) - \varGamma_{1,1} (R) \right) 
\\
&\qquad \qquad \qquad \times \left( \tfrac{1}{5} {\mathcal F} (3,1,0) - 6 {\mathcal F} (2,1,0) + 23 \, {\mathcal F} (1,1,0) + 432 \right) \Biggr]_{R=1/\sqrt{2}}
\\
&= - \frac{1}{8\pi}\, \frac{\partial}{\partial R} \Biggl[ \frac{1}{R} \Bigl( \tfrac{1}{5}\, {\mathcal I}_{1,1} ( 2 R; 3,1) -6 \, {\mathcal I}_{1,1} (2 R ; 2,1) +23 \,{\mathcal I}_{1,1} (2 R ; 1,1)  
\\
&\qquad \quad  - \left(\tfrac{1}{5}\, {\mathcal I}_{1,1} ( R ; 3,1) -6 \, {\mathcal I}_{1,1} ( R ; 2,1 ) +23 \,{\mathcal I}_{1,1} (R;1,1 ) \right) +144\,\pi ( R - \tfrac{1}{2 }\, R^{-1})\Bigr) \Biggr] _{R=1/\sqrt{2}}
\\
&= - 20 \, \sqrt{2}\,.
\end{split}
\end{equation}

\subsection*{Acknowledgements}

We are grateful to S. Hohenegger,  J. Manschot, S. Murthy for interesting comments and discussions. 
B. P. is grateful to R. Bruggeman, 
J. H. Bruinier, W. Pribitkin and \'A. T\'oth for helpful remarks during
the Krupp Symposium on Modular Forms, Mock Theta Functions and Applications
 in K\"oln, Feb 27 - March 1, 2012. 
 C.A and I.F. would like to thank the TH Unit at CERN for hospitality while this project was in progress. This work was partially supported by the European ERC Advanced Grant no. 
226455 ``Supersymmetry, Quantum Gravity and Gauge Fields'' (SUPERFIELDS) and by the Italian MIUR-PRIN 
contract 2009KHZKRX-007 ``Symmetries of the Universe and of the Fundamental Interactions''.

\appendix

\section{Notations and useful identities \label{sec_not}}

In this appendix we collect various definitions, notations and formulae used in the text.

\subsection{Operators acting on modular forms \label{sec_notmod}}

The hyperbolic Laplacian acts on modular forms of weight $w$ via\footnote{Our Laplacian is related to the one used {\em e.g} in \cite{1154.11015} 
via $\Delta_w=-\tfrac12 \Delta_{w;\rm BO}-\tfrac{w}{2}$.}
\be
\label{Deltaw}
\Delta_w = 2 \, \tau_2^2\, \partial_{\bar\tau}\, \left( \partial_\tau-\frac{\I w}{2\tau_2} \right)\ .
\end{equation}
We denote by $\cH(s,w)$ the eigenspace of $\Delta_w$ with eigenvalue
$ \frac12 s(s-1) -\frac18 w(w+2)$, in the space of real analytic functions 
of modular weight $w$  under $\varGamma ={\rm SL} (2, \mathbb{Z})$. 
The raising and lowering operators $D_w$, $\bar D_w$ defined by
\be
\label{modularderiv}
D_w=\frac{\I}{\pi} \left( \partial_\tau-\frac{\I w}{2\tau_2} \right) \ ,\qquad
\bar D_w=-\I \pi\, \tau_2^2 \partial_{\bar\tau}\,,
\ee
map $\cH(s,w)$ to $\cH(s ,w\pm 2)$,
\be
\cH(s ,w-2) \stackrel{\bar D_w}{\longleftarrow} \cH(s, w)
\stackrel{D_w}{\longrightarrow} \cH(s, w+2) \,,
\ee
and satisfy the commutation identity
\be
D_{w-2} \cdot \bar D_w - \bar D_{w+2}\cdot D_w = \frac{w}{4} \, .
\ee
The operator $D_w$ (and of course, $\bar D_w$) satisfies the Leibniz rule
\be
D_{w+w'}\, (f_w\, f_{w'}) = (D_w \, f_w)\, f_{w'} + f_w \, (D_{w'}\, f_{w'})\,,
\label{leibniz}
\ee
where $f_w$ is  a modular form of weight $w$. We denote by $D_w^r \, f_w$ (or simply $D^r f$) 
the iterated
derivative  $ D_{w+2r-2}\cdot \ldots \cdot D_{w+2}\cdot D_w\cdot f_w$, a modular form
of weight $w+2r$. One has
\be
\label{iteratedD}
D_w^r=  \left( \frac{\I}{\pi}\right)^r \, \sum_{j=0}^{r} \frac{r!}{j!\, (r-j)!}\, \frac{\varGamma(w+r)}{\varGamma(w+j)}\, 
(2\I\tau_2)^{j-r}\,\partial^j_\tau \ .
\ee
For $w\leq 0$, the operator $D^{1-w}_w$ simplifies to $(\I/\pi)^{1-w} \partial^{1-w}_\tau$ (Bol's identity), and is  
known in the physics literature as the Farey transform \cite{Dijkgraaf:2000fq}.

The Hecke operators $T_\kappa$ are defined by 
\be
\label{defHecke}
(T_\kappa \cdot \varPhi)(\tau)=\sum_{\substack{a,d>0 \\ad=\kappa}}\, \sum_{b\, {\rm mod}\, d} 
d^{-w}\, \varPhi\left( \frac{a\tau+b}{d}\right)\,,
\ee
and satisfy the commutative algebra 
\be
\label{HeckeAlg}
T_\kappa\, T_{\kappa'} = \sum_{d|(\kappa,\kappa')}\, d^{1-w}\,T_{\kappa \kappa'/d^2}\,,
\ee
If $\varPhi=\sum_{n\in\IZ} \varPhi(n,\tau_2)\, e^{2\pi n \tau_1}$ is a modular form
of weight $w$, then the Fourier coefficients of $T_\kappa\cdot \varPhi$ are
\be
\label{Heckemodes}
(T_\kappa \cdot \varPhi)(n,\tau_2)=\kappa^{1-w}\, \sum_{d|(n,\kappa)} 
d^{w-1}\, \varPhi\left( n\kappa/d^2, d^2 \tau_2/\kappa\right)\,.
\ee

The generators of the ring of holomorphic modular forms are the normalised 
Eisenstein series 
\be
E_4 = 1+240\, \sum_{n=0}^{\infty} \frac{n^3 q^n}{1-q^n}\qquad {\rm and}\qquad
E_6 = 1-504\, \sum_{n=0}^{\infty} \frac{n^5 q^n}{1-q^n}\ ,
\ee
with modular weight 4 and 6, respectively. The discriminant function
is the weight 12 cusp form 
\be
\varDelta = q \, \prod_{n=1}^{\infty} (1-q^n)^{24} = \tfrac{1}{1728}(E_4^3-E_6^2)\,.
\ee
The generators of the ring of weak holomorphic modular forms are 
$E_4, E_6$ and $1/\varDelta$. The modular $j$-invariant is the unique 
weak holomorphic modular form of weight zero with $j=1/q+\cO(q)$,
\be
j = \frac{E_4^3}{ \varDelta} -744 =\frac{E_6^2}{\varDelta}+984\ .
\ee
The ring of weak almost  holomorphic modular forms
is obtained by adding to $E_4, E_6,1/\varDelta$ the almost holomorphic Eisenstein series
\be
\label{defE2h}
\hat E_2 = E_2 - \frac{3}{\pi\tau_2} 
= 1-24\, \sum_{n=0}^{\infty} \frac{n\, q^n}{1-q^n}- \frac{3}{\pi\tau_2} \ .
\ee
Under the raising operator $D_w$ one has
\be
D \hat E_2 = \tfrac{1}{6}(E_4- \hat E_2^2 )\,,\quad 
D E_4 = \tfrac{2}{3}(E_6 - \hat E_2 E_4) \,,\quad 
D E_6 = E_4^2 - \hat E_2 E_6\,,\quad 
D (1/\varDelta) = 2 \hat E_2/\varDelta\,,
\ee
where, for simplicity, we have left implicit the specification of the weight in $D$.
Using the Leibniz rule \eqref{leibniz}, this allows to compute the action of $D$ on 
any weak almost  holomorphic modular form. 

Finally, the operator $T_\kappa$  maps the weak holomorphic modular form  $\varPhi=1/q+\cO(q)$ to $T_\kappa \varPhi=1/q^\kappa+\cO(q)$.

\subsection{Whittaker and hypergeometric functions} 
\label{WhittakerApp}

Whittaker functions and hypergeometric functions, more in general, are central in the analysis of the Niebur-Poincar\'e series and the evaluation of one-loop modular integrals. We summarise here their definitions and some of their main properties.

Whittaker functions are solutions of the second-order differential equation
\begin{equation}
u'' + \left( -\frac{1}{4} + \frac{\lambda}{z} + \frac{\frac{1}{4}-\mu^2}{z^2} \right) \, u=0\,.
\label{Whittakerde}
\end{equation}
For $2 \mu$ not integer the two independent solutions are given by the Whittaker $M$-functions
\begin{equation}
\label{M1F1}
M_{\lambda, \pm\mu}(z)=e^{-z/2}\, z^{\pm\mu+\tfrac12}\, 
_1 F_1\left(\pm\mu-\lambda+\tfrac12;1\pm2\mu ; z\right)\, ,
\end{equation}
and are expressed in terms of the confluent hypergeometric function 
\begin{equation}
{}_1 F_1 (a;b;z) = \frac{\varGamma (b)}{\varGamma (a)}\,\sum_{n=0}^\infty\frac{\varGamma (a+n)}{\varGamma (b+n)}\, \frac{z^n}{n!}\,.
\end{equation}
When $2 \mu$ is an integer, however, the second solution is not defined anymore, and thus it is useful to introduce a second Whittaker function defined by
\begin{equation}
W_{\lambda, \mu} (z) = - \frac{1}{2\pi\I}\varGamma (\lambda +\tfrac{1}{2}-\mu ) \,e^{-z/2} \, z^\lambda\, \int_\infty^{(0+)} (-t)^{-\lambda-\frac{1}{2}+\mu}\, \left( 1+\frac{t}{z} \right)^{\lambda -\frac{1}{2} +\mu}\, e^{-t}\, dt
\,,
\end{equation}
where $|{\rm arg} (-t)|\le \pi$, and the contour does not contain the point $t=-z$ and circles the origin counter-clockwise. The functions $W_{\lambda,\mu}(z)$ and $W_{-\lambda ,\mu }(-z)$ are two independent solutions of the differential equation \eqref{Whittakerde}, 
behaving as $(\pm z)^{\pm\lambda}
e^{\mp z/2}$ as $z\to +\infty$. As a result, the function $\cW_{s,w}$ defined in 
\eqref{defW} is exponentially suppressed as $y\to \pm \infty$,
\begin{equation}
\cW_{s,w}(y) \sim |4\pi y|^{\tfrac{w}{2}(\sgn(y)-1)}\, e^{-2\pi|y|} \qquad {\rm as}\quad |y|\to \infty \,.
\end{equation}
The Whittaker $M$-function can then be expressed as the linear combination
\begin{equation}
\label{WtoM}
M_{\lambda,\mu}(z) = \frac{\varGamma(2\mu+1)}{\varGamma(\mu-\lambda+\tfrac12)}
e^{\I\pi\lambda}\, W_{-\lambda,\mu}(e^{\I\pi}z)+
\frac{\varGamma(2\mu+1)}{\varGamma(\mu+\lambda+\tfrac12)}
e^{\I\pi(\lambda-\mu-\frac12)}\, W_{\lambda,\mu}(z) \,.
\end{equation}
Using the symmetry of the $W$-functions, $W_{\lambda , \mu} (z) = W_{\lambda , -\mu} (z)$, one can invert the previous relation and write
\begin{equation}
\label{MtoW}
W_{\lambda , \mu} (z) = \frac{\varGamma (-2\mu )}{\varGamma (\frac{1}{2} -\mu-\lambda)} \, M_{\lambda , \mu} (z) + \frac{\varGamma (2\mu)}{\varGamma (\frac{1}{2}+\mu-\lambda ) } \, M_{\lambda , -\mu } (z)\,.
\end{equation}
This implies that the functions ${\mathcal M}_{s,w}$ and ${\mathcal W}_{s,w}$ obey
\begin{equation}
{\mathcal W}_{s,w} (y) = \frac{\varGamma (1-2s)}{\varGamma (1-s-\frac{w}{2}\,{\rm sgn} (y))}\, {\mathcal M}_{s,w} (y) + \frac{\varGamma (2s-1)}{\varGamma (s-\frac{w}{2}\, {\rm sgn} (y))}\, {\mathcal M}_{1-s,w} (y)\,.
\end{equation}

For special values of $\lambda$ and $\mu$, the Whittaker functions reduce to elementary functions or to other special functions. This derives from the properties of the hypergeometric functions, for instance
\begin{equation}
\begin{split}
_1 F_1(a;a;z) &=e^z \,,
 \\
_1 F_1(1,a,z) &=(a-1)\,z^{1-a} \, e^z\, \gamma(a-1,z)\,,
\\
_1 F_1(a,a+1,z) &= a (-z)^{-a} \, \gamma(a,-z) \,,
\\
 _1 F_1(a; 2 a;z) &=  e^{z/2} \, \left( \tfrac{1}{4}\, z\right)^{\frac{1}{2}-a} \varGamma (a+\tfrac{1}{2} )
\, I_{a-\frac{1}{2}} (z/2)
 \,,
\end{split}
\label{prop1F1}
\end{equation}
where
\begin{equation}
\gamma (a,z ) = \int_0^z e^{-t}\, t^{a-1} dt =\varGamma (a) - \varGamma (a,z)
\end{equation}
is the incomplete Gamma function. The Bessel functions $I_\nu, J_\nu, K_\nu$ are defined by  
\begin{equation}
\label{BesselIJ}
I_\nu (z) = \sum_{m=0}^\infty \frac{(z/2)^{2m+\nu}}{m!\, \varGamma (m+\nu+1)} = \I^{-\nu} \, J_{\nu} (\I z)
\end{equation}
\begin{equation}
K_\nu (z) = \frac{\pi}{2}\, \frac{I_{-\nu} (z) - I_\nu (z)}{\sin \, \pi\nu}\,,
\end{equation}
Using \eqref{M1F1}, \eqref{MtoW}, \eqref{curlyM}, \eqref{defW}, 
 one finds, for $y>0$, 
\begin{equation}
\begin{split}
{\mathcal M}_{s,0} (\pm y) &= 2^{2s-1}\, \varGamma (s+\tfrac{1}{2})\, (4\pi |y|)^{\frac{1}{2}}\, I_{s-\frac{1}{2}} (2 \pi |y|) \,,
\\
{\mathcal W}_{s,0} (\pm y) &= 2\, |y|^\frac{1}{2} \, K_{s-\frac{1}{2}} (2\pi |y|)\,,
\end{split}
\label{Whittakers0}
\end{equation}
\begin{equation}
\label{limM1}
\begin{split}
\cM_{\tfrac{w}{2},w}(-y) &= e^{2\pi y}\ ,
\\
\cW_{\tfrac{w}{2},w}(-y) &=\varGamma(1-w,4\pi y)\,  e^{2\pi y}\,,\quad
\\
\cW_{\tfrac{w}{2},w}(y) &= e^{-2\pi y}\,,
\end{split}
\end{equation}
\begin{equation}
\label{limM2}
\begin{split}
\cM_{-\tfrac{w}{2},w}(-y) &=(4\pi y)^{-w}\, e^{-2\pi y}\,,\quad
\\
\cW_{-\tfrac{w}{2},w}(-y) &= (4\pi y)^{-w}\, e^{-2\pi y}\,, \quad
\end{split}
\end{equation}
\begin{equation}
\label{limM3}
\begin{split}
\cM_{1-\tfrac{w}{2},w}(-y) &=(-1)^{(w+1)} \, (1-w)\, 
\gamma(1-w,4\pi y)\,  e^{2\pi y}\,,
\\
{\mathcal W}_{1-\frac{w}{2},w} (-y) &= \varGamma (1-w,4\pi y)\, e^{2\pi y}\,
\ ,\quad 
\\
{\mathcal W}_{1-\frac{w}{2},w} (y) &= e^{-2\pi y}\,.
\end{split}
\end{equation}

For integer values of the arguments, the confluent hypergeometric function, and thus the Whittaker functions, take a particularly simple expression 
\begin{equation}
\begin{split}
_1 F_1(n+1,a,z) &= \frac{\varGamma(a)}{n!} \, \frac{\de^ n}{\de z^n}
\left[ z^{n+1-a}\, \left( e^z - \sum_{k=0}^{a-2} \frac{z^k}{k!} \right) \right]
\\
&= \varGamma (a) \, z^{1-a} \left[ e^z\, L_n^{(1-a)} (-z) - L^{(1-a)}_{a-2-n} (z)\right]\,,
\end{split}
\end{equation}
where
\begin{equation}
\label{defLag}
L_n^{(k)}(x) = \frac{x^{-k} \, e^x}{n!}\,  \frac{\de^ n}{\de x^n}\, \left[ x^{n+k}\, e^{-x}\right]
\end{equation}
are the associated Laguerre polynomials. 

As a result, the seed function that enters in the definition of the Niebur-Poincar\'e series involves only a finite number of terms when $s=1-\frac{w}{2}+n$, and reads, for $y>0$, 
\begin{equation}
\begin{split}
\cM_{1-\tfrac{w}{2}+n,w}(-y) =& (4\pi y)^{1-w+n} e^{-2\pi y} \frac{(2n+1-w)!}{n!}
\\
&\times
 \frac{\de^ n}{\de (4\pi y)^n}\,  
\left[(4\pi y)^{w-n-1} \left( e^{4\pi y} -
\sum_{k=0}^{2n-w} \frac{(4\pi y)^{k}}{k!} \right) \right]
\\
=&
\varGamma (2n+2-w) \, (4\pi y)^{-n} 
\\
&\times \left[ e^{2\pi y} \, L_n^{(-1-2n+w)} (-4\pi y) - e^{-2\pi y}\, L_{n-w}^{(-1-2n+w)} (4\pi y)\right]\,.
\end{split}
\label{Mn}
\end{equation}
Similarly,
\be
\begin{split}
\cW_{1-\tfrac{w}{2}+n,w}(y) =& (-1)^n\, n!\, (4\pi y)^{-n}\, e^{-2\pi y}\, L_n^{(-1-2n+w)} (4\pi y) \,,
\\
\cW_{1-\tfrac{w}{2}+n,w}(-y) =& (-1)^{n-w}\, \varGamma (n-w+1)\, (4\pi y)^{-n}\, e^{-2\pi y}\, L_{n-w}^{(-1-2n+w)} (4\pi y)\,.
\end{split}
\label{Wn}
\ee

The modular derivatives \eqref{modularderiv} have a simple action on  Whittaker functions, 
\begin{equation}
\label{DwM}
\begin{split}
D_w \cdot \left[ \cM_{s,w}(-\kappa\tau_2)  \, e^{-2\pi\I \kappa\tau_1}\right] &=  2\kappa (s+\tfrac{w}{2})\,  \cM_{s,w+2}(-\kappa\tau_2)  \, e^{-2\pi\I \kappa\tau_1}\,,
\\
\bar D_w \cdot \left[ \cM_{s,w}(-\kappa\tau_2)  \, e^{-2\pi\I \kappa\tau_1}\right] &= \frac{1}{8\kappa} (s-\tfrac{w}{2})\,   \cM_{s,w-2}(-\kappa\tau_2)  \, e^{-2\pi\I \kappa\tau_1}\,,
\end{split}
\end{equation}
\begin{equation}
\label{DwW}
\begin{split}
D_w\cdot \left[ \cW_{s,w}(n\tau_2)  \, e^{2\pi\I n\tau_1}\right] &= 
\cW_{s,w+2}(n\tau_2)\, e^{2\pi\I n\tau_1} \times 
\begin{cases} 
-2 n \ , & n>0 \,,
\\
2 n \, (s+\tfrac{w}{2})(s-\tfrac{w}{2}-1)\ , & n<0 \,,
\end{cases} 
\\
\bar D_w \cdot \left[ \cW_{s,w}(n\tau_2)  \, e^{2\pi\I n\tau_1}\right] &= 
\cW_{s,w-2}(n\tau_2)\, e^{2\pi\I n\tau_1} \times
\begin{cases}  
\frac{1}{8 n} & n<0\,,
\\
- \frac{1}{8 n}\, 
(s-\tfrac{w}{2})\,(s+\tfrac{w}{2}-1)\, & n>0\,.
\end{cases}
\end{split}
\end{equation}

\subsection{Kloosterman-Selberg zeta function\label{Kloosterman}}

The Kloosterman-Selberg zeta function entering in the expression \eqref{FourierFBO}
of the Fourier coefficients of the Niebur-Poincar\'e series is defined as
\cite{1006.11024}
\be
\label{Ziv}
\cZ_s (a,b) = 
\frac{1}{2\sqrt{|ab|}}\, \sum_{c>0} \frac{S (a,b;c)}{c} \, \times
\begin{cases}
J_{2s-1} \left( \frac{4\pi}{c} \sqrt{a b}\right)  & \mbox{if}\quad a\, b>0\,,
\\
I_{2s-1} \left( \frac{4\pi}{c} \sqrt{-a b}\right) & \mbox{if}\quad a\, b<0 \,,
\end{cases}
\ee
where $I_s(x)$ and $J_s(x)$ are the Bessel $I$ and $J$ functions, and 
$S(a,b;c)$ are  the classical Kloosterman sums 
for the modular group $\varGamma = {\rm SL} (2,\mathbb{Z})$,
\be
S (a,b;c) = \sum_{d\in (\mathbb{Z}/c\mathbb{Z})^*} \exp\left[ \frac{2\pi \I}{c} (a \, d+ b\, d^{-1})\right]\,.
\ee
Here $a$, $b$ and $c$ are integers, and $d^{-1}$ is the inverse of $d$ mod $c$. $S (a,b;c)$ is clearly symmetric under the exchange of $a$ and $b$. Less evidently, it satisfies the Selberg identity
\be 
\label{SelbergId}
S (a,b;c) = \sum_{d | {\rm gcd} (a,b,c)} d\, S ( ab/d^2,1;c/d)\,.
\ee
In the special case $a\neq 0$, $b=0$, the Kloosterman sum reduces to the Ramanujan sum
\be
S(a,0;c) = S(0,a;c) = \sum_{d\in (\mathbb{Z}/c\mathbb{Z})^*} \exp \left( \frac{2\pi \I}{c} a\, d \right) = \sum_{d | {\rm gcd} (c,a)}d\,  \mu (c/d)\,,
\ee
with $\mu (n)$ the M\"obius function. For $a=b=0$, $S (a,b;c)$ reduces instead to the 
Euler totient function $\phi(c)$, and one can verify that
\begin{equation}
\sum_{c>0}\frac{S (0,0;c)}{c^{2s}} = \frac{\zeta (2s-1)}{\zeta (2s)}\,,
\qquad
\sum_{c>0}\frac{S (0,\pm \kappa;c )}{c^{2s}} = \frac{\sigma_{1-2s} (\kappa )}{\zeta (2s)}\qquad (\kappa \not=0 )\,,
\end{equation}
with $\sigma_x (n)$ the divisor function.
 Under complex conjugation, $\cZ_s (a,b)$  transforms as 
\be
\label{Zcj}
\overline{\cZ_s (a,b)} = \cZ_{\bar s} (-a,-b)\ .
\ee
The Kloosterman-Selberg zeta function defined in \eqref{Ziv} is related to the 
zeta function 
\be
Z(a,b;s) \equiv \sum_{c>0} \frac{S (a,b;c)}{c^{2s}}
\ee
originally considered in
\cite{0142.33903} and used in \cite{Angelantonj:2011br} via
\begin{equation}
{\mathcal Z}_s (a,b) =\pi \, (4\pi^2 |a\, b|)^{s-1} \sum_{m=0}^\infty \frac{(-4\pi^2 a \, b)^m}{m!\, \varGamma (2s+m)}\, Z(a,b;s+m)\,.
\end{equation}

\section{Selberg-Poincar\'e series vs. Niebur-Poincar\'e series \label{sec_Selberg}}

In this section, we briefly discuss the relation between the Niebur-Poincar\'e series \eqref{Fskw}
and the Selberg-Poincar\'e series \eqref{defEgen}, considered in our previous work \cite{Angelantonj:2011br} in the special case $w=0$, as well as the relation between
the BPS-state sum \eqref{int2F1} and the ``shifted constrained Epstein zeta series''
considered in \cite{Angelantonj:2011br}.

Comparing  the differential equations \eqref{laplFskw} and  \eqref{laplEskw}, it is easily
seen that a set of solutions of one can be converted into a set of solutions of the other
by considering the linear combinations \cite{0543.10020,0741.11024}
\be
\label{FtoE}
\begin{split}
\cF(s,\kappa,w)=&\sum_{m\geq 0} a(s,\kappa,w,m)\, E(s+m,\kappa,w)\,,
\\
E(s,\kappa,w)=&\sum_{m\geq 0} b(s,\kappa,w,m)\, \cF(s+m,\kappa,w)\,,
\end{split}
\ee
such that the coefficients satisfy the recursion relations
\be
\frac{a(s,\kappa,w,m+1)}{a(s,\kappa,w,m)}=-
\frac{4\pi\kappa (s+m-\tfrac{w}{2})}{(m+1)(m+2s)}\ ,\qquad
\frac{b(s,\kappa,w,m+1)}{b(s+1,\kappa,w,m)}=
\frac{4\pi\kappa (s-\tfrac{w}{2})}{(m+1)(m+2s)}\ .
\ee
Comparing also the constant term \eqref{Fzero} of the 
Niebur-Poincar\'e series and the constant term
\be
\label{Etilde00}
\tilde E_0(s,\kappa,w) = \sum_{m=0}^\infty
\frac{ 2^{2(1-s)}\, \pi\, \I^{-w}\,   (\pi\kappa)^m\,  \varGamma (2s + m -1)\, \sigma_{1-2s-2m}(\kappa)}
{m!\, \varGamma (\frac{w}{2} + s + m) \, \varGamma (s -\frac{w}{2})\, \zeta(2s+2m)}\, 
\tau_2^{1-s-m-\tfrac{w}{2}}\ ,
\ee
of the Selberg-Poincar\'e series, we find that the coefficients are given by 
\begin{equation}
\begin{split}
a(s,\kappa,w,m) &= (-1)^m\,
\frac{2^{2s+2m-w}\, (\pi\kappa)^{s-\tfrac{w}{2}+m}\, 
\varGamma(2s)\, \varGamma(s+m-\tfrac{w}{2})}
{m!\, \varGamma(2s+m)\, \varGamma(s-\tfrac{w}{2})} \,,
\\
b(s,\kappa,w,m) &= 
\frac{2^{w-2s}\, (\pi\kappa)^{-s+\tfrac{w}{2}}\,
\varGamma(2s+m-1)\, \varGamma(s+m-\tfrac{w}{2})}
{m!\, \varGamma(2s+2m-1)\, \varGamma(s-\tfrac{w}{2})} \,.
\end{split}
\end{equation}
In particular, in the limit $s\to \tfrac{w}{2}$ where the summand of the 
Selberg-Poincar\'e series \eqref{defEgen} becomes holomorphic, one finds
 $E(s,\kappa,w)=\cF(\tfrac{w}{2},\kappa,w)$ for $w\geq 2$, but
\be
\label{EtoF}
E(\tfrac{w}{2},\kappa,w)= \cF(\tfrac{w}{2},\kappa,w)+
\sum_{m=1}^{-\frac{w}{2}-1} b'_m \, {\rm Res}_{s=\frac{w}{2}+m} \, \cF(s,\kappa,w)+
\sum_{m=-\frac{w}{2}+1}^{1-w} b_m \, \cF(\tfrac{w}{2}+m,\kappa,w)
\ee
for $w\leq 0$, where $b_m\equiv \lim_{s\to\frac{w}{2}} b(s,\kappa,w,m)$ 
and $b'_m\equiv \lim_{s\to\frac{w}{2}}\frac{\de}{\de s}b(s,\kappa,w,m)$. 
In writing \eqref{EtoF}, we have assumed that the singularities 
of $\cF(s,\kappa,w)$ on the real $s$-axis can be read off from the constant term 
\eqref{Fzero}, namely that 
$\cF(s,\kappa,w)$ is regular at integer values of $s$ provided $s\geq 0$ or $s\leq -|w|/2$,
and has simple poles for integer values of $s$ such that $-\frac{|w|}{2} <s<0$. In particular, 
$E(\tfrac{w}{2},\kappa,w)$ receives a contribution (for $m=1-w$) proportional to 
the harmonic Maass form $\cF(1-\tfrac{w}{2},\kappa,w)$, which is the main object
of interest in the present work, but is contaminated by other Niebur-Poincar\'e
series lying outside the convergence domain $\Re(s)>1$. For example, for $w=0$, we find
\be
E(0,\kappa,0) = \cF(0,\kappa,0) + \tfrac12 \cF(1,\kappa,0) \,,
\ee
consistently with the identifications 
\be
E(0,\kappa,0) = T_\kappa\, j+12 \sigma(\kappa)\,,\quad 
\cF(1,\kappa,0) = T_\kappa\, j+24 \sigma(\kappa)\,,\quad
 \cF(0,\kappa,0) = \tfrac12 T_\kappa\, j\,.
\ee
Moreover, the relation \eqref{FtoE} between the Niebur-Poincar\'e and 
Selberg-Poincar\'e series implies a similar relation between the BPS-state 
sum \eqref{int2F1} and the ``shifted constrained Epstein zeta series''
\be
\label{defsEpstein}
\mathcal{E}^{d}_V (G,B,Y;s,\kappa) \equiv 2^s\,
\sum_{\rm BPS} (p_L^2+p_R^2-4\,\kappa)^{-s} =  \sum_{\rm BPS} ( p_{\rm R}^2 )^{-s} 
\ee
generalising the constructio in \cite{Angelantonj:2011br} to the case $k\not=0$. Namely, using the same techniques as in our previous paper, one may show that \eqref{defsEpstein} arises from the modular integral
\be
\label{intEskwd}
\lim_{\cT\to\infty}
\int_{\cF_\cT}\de\mu\, \varGamma_{d+k,d}(G,B,Y) \, E( s,\kappa,-\tfrac{k}{2}) 
= \frac{\varGamma (s+\frac{2d+k}{4}-1)}{\pi^{s+\frac{2d+k}{4}-1}} \, \mathcal{E}^{d}_V (G,B,Y;
s+\tfrac{2s+k}{4}-1,\kappa) \,.
\ee
Since the BPS-state sum ${\mathcal I}_{d+k,d} (s,\kappa )$ arises from the
limit  $\cT\to\infty$ of the integral \eqref{Irenorm1},
from \eqref{FtoE} we conclude that
\be
\mathcal{E}^{d}_V (G,B,Y;
s+\tfrac{2d+k}{4}-1,\kappa) =  \frac{\pi^{s+\frac{2d+k}{4}-1}} {\varGamma (s+\frac{2d+k}{4}-1)}
\sum_{m\geq 0} b(s,\kappa,w,m)\, {\mathcal I}_{d+k,d} (s+m,\kappa )\,,
\ee
for large $\Re(s)$.



\begin{thebibliography}{10}

\bibitem{Kiritsis:1999ss}
E.~Kiritsis, ``{Duality and instantons in string theory},''
\href{http://arxiv.org/abs/hep-th/9906018}{{\tt arXiv:hep-th/9906018}}.

\bibitem{Lerche:1987qk}
W.~Lerche, B.~Nilsson, A.~Schellekens, and N.~Warner, ``{Anomaly cancelling
  terms from the elliptic genus},''
\href{http://dx.doi.org/10.1016/0550-3213(88)90468-3}{{\em Nucl.Phys.} {\bf
  B299} (1988)  91}.

\bibitem{MR656029}
D.~Zagier, ``The {R}ankin-{S}elberg method for automorphic functions which are
  not of rapid decay,'' {\em J. Fac. Sci. Univ. Tokyo Sect. IA Math.} {\bf 28}
  (1981) no.~3, 415--437 (1982).

\bibitem{Kutasov:1990sv}
D.~Kutasov and N.~Seiberg, ``{Number of degrees of freedom, density of states
  and tachyons in string theory and CFT},''
\href{http://dx.doi.org/10.1016/0550-3213(91)90426-X}{{\em Nucl.Phys.} {\bf
  B358} (1991)  600--618}.

\bibitem{Angelantonj:2010ic}
C.~Angelantonj, M.~Cardella, S.~Elitzur, and E.~Rabinovici, ``{Vacuum
  stability, string density of states and the Riemann zeta function},''
  \href{http://dx.doi.org/10.1007/JHEP02(2011)024}{{\em JHEP} {\bf 1102} (2011)
   024},
\href{http://arxiv.org/abs/1012.5091}{{\tt arXiv:1012.5091 [hep-th]}}.

\bibitem{Cardella:2010bq}
M.~A. Cardella, ``{Error Estimates in Horocycle Averages Asymptotics:
  Challenges from String Theory},''
\href{http://arxiv.org/abs/1012.2754}{{\tt arXiv:1012.2754 [math.NT]}}.

\bibitem{Cardella:2008nz}
M.~Cardella, ``{A novel method for computing torus amplitudes for
  $\mathbb{Z}_{N}$ orbifolds without the unfolding technique},''
  \href{http://dx.doi.org/10.1088/1126-6708/2009/05/010}{{\em JHEP} {\bf 05}
  (2009)  010},
\href{http://arxiv.org/abs/0812.1549}{{\tt arXiv:0812.1549 [hep-th]}}.

\bibitem{McClain:1986id}
B.~McClain and B.~D.~B. Roth, ``{Modular invariance for interacting bosonic
  strings at finite temperature},''
\href{http://dx.doi.org/10.1007/BF01219073}{{\em Commun.Math.Phys.} {\bf 111}
  (1987)  539}.

\bibitem{O'Brien:1987pn}
K.~O'Brien and C.~Tan, ``{Modular Invariance of Thermopartition Function and
  Global Phase Structure of Heterotic String},''
  \href{http://dx.doi.org/10.1103/PhysRevD.36.1184}{{\em Phys.Rev.} {\bf D36}
  (1987)  1184}.
Preliminary Draft.

\bibitem{Dixon:1990pc}
L.~J. Dixon, V.~Kaplunovsky, and J.~Louis, ``Moduli dependence of string loop
  corrections to gauge coupling constants,''
{\em Nucl. Phys.} {\bf B355} (1991)  649--688.

\bibitem{Mayr:1993mq}
P.~Mayr and S.~Stieberger, ``{Threshold corrections to gauge couplings in
  orbifold compactifications},''
  \href{http://dx.doi.org/10.1016/0550-3213(93)90096-8}{{\em Nucl. Phys.} {\bf
  B407} (1993)  725--748},
\href{http://arxiv.org/abs/hep-th/9303017}{{\tt arXiv:hep-th/9303017}}.

\bibitem{Bachas:1997mc}
C.~Bachas, C.~Fabre, E.~Kiritsis, N.~A. Obers, and P.~Vanhove,
  ``{Heterotic/type-I duality and D-brane instantons},''
  \href{http://dx.doi.org/10.1016/S0550-3213(97)00639-1}{{\em Nucl. Phys.} {\bf
  B509} (1998)  33--52},
\href{http://arxiv.org/abs/hep-th/9707126}{{\tt arXiv:hep-th/9707126}}.

\bibitem{stst1}
W. Lerche and S. Stieberger, 
``{Prepotential, mirror map and F theory on K3},''
{\em Adv. Theor. Math. Phys.} {\bf 2} (1998) 1105-1140, {\tt arXiv:hep-th/9804176}.

\bibitem{stst2}
K.~Foerger and S.~Stieberger,
``{Higher derivative couplings and heterotic type I duality in eight-dimensions},''
{\em Nucl. Phys.} {\bf B559} (1999) 277-300, {\tt arXiv:hep-th/9901020}.

\bibitem{Kiritsis:1997hf}
E.~Kiritsis and N.~A. Obers, ``{Heterotic/type-I duality in D < 10 dimensions,
  threshold corrections and D-instantons},'' {\em JHEP} {\bf 10} (1997)  004,
\href{http://arxiv.org/abs/hep-th/9709058}{{\tt arXiv:hep-th/9709058}}.

\bibitem{Kiritsis:1997em}
E.~Kiritsis and B.~Pioline, ``{On $R^4$ threshold corrections in type IIB
  string theory and $(p,q)$ string instantons},''
  \href{http://dx.doi.org/10.1016/S0550-3213(97)00645-7}{{\em Nucl. Phys.} {\bf
  B508} (1997)  509--534},
\href{http://arxiv.org/abs/hep-th/9707018}{{\tt arXiv:hep-th/9707018}}.

\bibitem{Marino:1998pg}
M.~Marino and G.~W. Moore, ``{Counting higher genus curves in a Calabi-Yau
  manifold},'' \href{http://dx.doi.org/10.1016/S0550-3213(98)00847-5}{{\em
  Nucl. Phys.} {\bf B543} (1999)  592--614},
\href{http://arxiv.org/abs/hep-th/9808131}{{\tt arXiv:hep-th/9808131}}.


\bibitem{Harvey:1995fq}
J.~A. Harvey and G.~W. Moore, ``{Algebras, BPS States, and Strings},'' {\em
  Nucl. Phys.} {\bf B463} (1996)  315--368,
\href{http://arxiv.org/abs/hep-th/9510182}{{\tt hep-th/9510182}}.

\bibitem{Harvey:1996gc}
J.~A. Harvey and G.~W. Moore, ``On the algebras of {BPS} states,'' {\em Commun.
  Math. Phys.} {\bf 197} (1998)  489--519,
\href{http://arxiv.org/abs/hep-th/9609017}{{\tt hep-th/9609017}}.

\bibitem{Angelantonj:2011br}
C.~Angelantonj, I.~Florakis, and B.~Pioline, ``{A new look at one-loop
  integrals in string theory},'' to appear in {\em Commun. Num. Theor. Phys.}, 
\href{http://arxiv.org/abs/1110.5318}{{\tt arXiv:1110.5318 [hep-th]}}.

\bibitem{Obers:1999um}
N.~A. Obers and B.~Pioline, ``{Eisenstein series and string thresholds},''
  \href{http://dx.doi.org/10.1007/s002200050022}{{\em Commun. Math. Phys.} {\bf
  209} (2000)  275--324},
\href{http://arxiv.org/abs/hep-th/9903113}{{\tt arXiv:hep-th/9903113}}.

\bibitem{Ooguri:1991fp}
  H.~Ooguri and C.~Vafa,
 ``{Geometry of N=2 strings},''
{\em  Nucl.\ Phys.}\ {\bf B361} (1991) 469.

\bibitem{Ferrara:1991uz}
  S.~Ferrara, C.~Kounnas, D.~Lust and F.~Zwirner,
 ``{Duality invariant partition functions and automorphic superpotentials for (2,2) string compactifications},''
{\em  Nucl.\ Phys.}\  {\bf B365} (1991) 431.

\bibitem{LopesCardoso:1994ik}
  G.~Lopes Cardoso, D.~Lust and T.~Mohaupt,
  ``Threshold corrections and symmetry enhancement in string compactifications,''
  {\em Nucl.\ Phys.}\ {\bf B450} (1995) 115,
  {\tt hep-th/9412209}.

\bibitem{LopesCardoso:1995qa}
  G.~Lopes Cardoso, G.~Curio, D.~Lust, T.~Mohaupt and S.~-J.~Rey,
 ``BPS spectra and nonperturbative gravitational couplings in N=2, N=4 supersymmetric string theories,''
 {\em Nucl.\ Phys.}\ {\bf B464} (1996) 18, 
{\tt hep-th/9512129}.
  
\bibitem{Lerche:1999ju}
  W.~Lerche and S.~Stieberger,
 ``1/4 BPS states and nonperturbative couplings in N=4 string theories,''
  {\em Adv.\ Theor.\ Math.\ Phys.}\  {\bf 3} (1999) 1539,
{\tt hep-th/9907133}.

\bibitem{Vigneras}
M.-F.~Vign\'eras, ``S\'eries Th\'eta des formes quadratiques ind\'efinies'',
International Summer School on Modular Functions, Bonn, 1976.

\bibitem{0306.30023}
D.~Niebur, ``{Construction of automorphic forms and integrals},''
  \href{http://dx.doi.org/10.2307/1997003}{{\em Trans. Am. Math. Soc.} {\bf
  191} (1974)  373--385}.

\bibitem{0695.10021}
M.~I. Knopp, ``{Rademacher on $J(\tau)$, Poincar\'e series of nonpositive
  weights and the Eichler cohomology},'' {\em Notices Am. Math. Soc.} {\bf 37}
  (1990) no.~4, 385--393.

\bibitem{Manschot:2007ha}
J.~Manschot and G.~W. Moore, ``{A Modern Fareytail},'' {\em Commun. Num. Theor.
  Phys.} {\bf 4} (2010)  103--159,
\href{http://arxiv.org/abs/0712.0573}{{\tt arXiv:0712.0573 [hep-th]}}.

\bibitem{0142.33903}
A.~Selberg, ``{On the estimation of Fourier coefficients of modular forms},''
  {\em Proc. Sympos. Pure Math.} {\bf 8} (1965)  1--15.

\bibitem{0288.10010}
D.~Niebur, ``{A class of nonanalytic automorphic functions},'' {\em Nagoya
  Math. J.} {\bf 52} (1973)  133--145.

\bibitem{0543.10020}
D.~A. Hejhal, {\em {The Selberg trace formula for $PSL(2,{\IR})$, Vol. 2.}}
\newblock {Lecture Notes in Math., Springer, no. 1001 (1983)}.

\bibitem{1004.11021}
J.~H. Bruinier, {\em {Borcherds products on $O(2,l)$ and Chern classes of
  Heegner divisors.}}
\newblock {Berlin: Springer}, 2002.

\bibitem{BruinierOno}
J.~Bruinier and K.~Ono, ``{Heegner divisors, $L$-functions and harmonic weak
  Maass forms},'' \href{http://dx.doi.org/10.4007/annals.2010.172.2135}{{\em
  Ann. Math. (2)} {\bf 172} (2010) no.~3, 2135--2181}.

\bibitem{1154.11015}
K.~Bringmann and K.~Ono, ``{Arithmetic properties of coefficients of
  half-integral weight Maass-Poincar\'e series},''
  \href{http://dx.doi.org/10.1007/s00208-006-0048-0}{{\em Math. Ann.} {\bf 337}
  (2007) no.~3, 591--612}.

\bibitem{OnoMockDelta}
K.~Ono, \href{http://dx.doi.org/10.1515/9783110208504.141}{``{A mock theta
  function for the delta-function},''}
\newblock {Berlin: Walter de Gruyter}, 2009.

\bibitem{0919.11036}
R.~E. Borcherds, ``{Automorphic forms with singularities on Grassmannians},''
  \href{http://dx.doi.org/10.1007/s002220050232}{{\em Invent. Math.} {\bf 132}
  (1998) no.~3, 491--562}.

\bibitem{0507.10029}
D.~Goldfeld and P.~Sarnak, ``{Sums of Kloosterman sums},''
  \href{http://dx.doi.org/10.1007/BF01389098}{{\em Invent. Math.} {\bf 71}
  (1983)  243--250}.

\bibitem{0352.30012}
J.~Fay, ``{Fourier coefficients of the resolvent for a Fuchsian group},''
  \href{http://dx.doi.org/10.1515/crll.1977.293-294.143}{{\em J. Reine Angew.
  Math.} {\bf 293/294} (1977)  143--203}.

\bibitem{1088.11030}
J.~H. Bruinier and J.~Funke, ``{On two geometric theta lifts.},''
  \href{http://dx.doi.org/10.1215/S0012-7094-04-12513-8}{{\em Duke Math. J.}
  {\bf 125} (2004) no.~1, 45--90}.

\bibitem{Dijkgraaf:2000fq}
R.~Dijkgraaf, J.~M. Maldacena, G.~W. Moore, and E.~P. Verlinde, ``{A black hole
  Farey tail},''
\href{http://arxiv.org/abs/hep-th/0005003}{{\tt arXiv:hep-th/0005003}}.

\bibitem{1006.11024}
H.~Iwaniec, {\em {Spectral methods of automorphic forms. 2nd ed.}}
\newblock {Providence, RI: American Mathematical Society (AMS); Madrid: Revista
  Matem\'atica Iberoamericana}, 2002.

\bibitem{0741.11024}
E.~Yoshida, ``{On Fourier coefficients of non-holomorphic Poincar\'e
  series},'' \href{http://dx.doi.org/10.2206/kyushumfs.45.1}{{\em Mem. Fac.
  Sci., Kyushu Univ., Ser. A} {\bf 45} (1991) no.~1, 1--17}.

\bibitem{luca} L.~Carlevaro E.~Dudas and D.~Israel, ``Gauge threshold corrections for $N =2$ heterotic local models with flux, and Mock modular forms,''
{\em to appear}.

\end{thebibliography}

\providecommand{\href}[2]{#2}\begingroup\raggedright\endgroup

\end{document}